\documentclass[useAMS,usenatbib]{mnras}

\usepackage{amsmath}
\usepackage{graphicx}
\usepackage{color}
\usepackage{hyperref}

\def\be{\begin{equation}}
\def\ee{\end{equation}}
\def\bea{\begin{eqnarray}}
\def\eea{\end{eqnarray}}
\def\nn{\nonumber}

\def\l{\left}
\def\r{\right}
\def\dt{\dot \tau}
\def\dg{d_\gamma}

\def\Dn{D_\nu}
\def\TP{[\Theta_0 + \Phi]}
\def\TS{[\Theta_0 - \Psi]}
\def\phig{\phi_{{\rm gr},\gamma}}
\def\phin{\phi_{{\rm gr},\nu}}

\def\Dl{D_\ell^{\rm TT}}
\def\Dle{D_\ell^{\rm EE}}
\def\Dlc{D_\ell^{\rm TE}}
\def\Cl{C_\ell^{\rm TT}}

\def\gtrsim{\ga}
\def\lesssim{\la}

\def \Planck {{\it Planck}\ }

\title[CMB Acoustic Peak Locations]{Cosmic Microwave Background Acoustic Peak Locations}

\author[Z. Pan et al.]{
	Z.~Pan,  	$^1$\thanks{Email: zhpan@ucdavis.edu}
        L.~Knox, 	$^1$\thanks{Email: lknox@ucdavis.edu}
	B.~Mulroe,	$^2$\thanks{Email: bmulroe@fordham.edu}
	A.~Narimani,	$^3$\thanks{Email: anariman@phas.ubc.ca}\\
$^1$Department of Physics, University of California, One Shields Avenue, Davis, CA, USA 95616\\
$^2$Department of Physics, Fordham University,  441 East Fordham Road, Bronx, NY, USA 10458\\
$^3$Department of Physics \& Astronomy, University of British Columbia, 6224 Agricultural Road, Vancouver, BC, V6T 1Z1 Canada}

\date{\today}

\begin{document}
\label{firstpage}

\maketitle

\begin{abstract}
The \Planck collaboration has measured the temperature and polarization of the
cosmic microwave background well enough to determine the locations of eight
peaks in the temperature (TT) power spectrum, five peaks in the polarization
(EE) power spectrum and twelve extrema in the cross (TE) power spectrum.
The relative locations of these extrema give a striking, and beautiful,
demonstration of what we expect from acoustic oscillations in the plasma;
e.g., that EE peaks fall half way between TT peaks. We expect this because
the temperature map is predominantly sourced by temperature variations in the
last scattering surface, while the polarization map is predominantly sourced by
gradients in the velocity field, and the harmonic oscillations have temperature
and velocity 90 degrees out of phase. However, there are large differences in
expectations for extrema locations from simple analytic models vs. numerical
calculations.  Here we quantitatively explore the origin of these differences
in gravitational potential transients, neutrino free-streaming, the breakdown
of tight coupling, the shape of the primordial power spectrum, details of the
geometric projection from three to two dimensions, and the thickness of the
last scattering surface. We also compare the peak locations determined from
\Planck measurements to expectations under the $\Lambda$CDM model. Taking into
account how the peak locations were determined, we find them to be in agreement.
\end{abstract}


\section{Introduction}
With the first release of \Planck polarization data \citep{PlanckXI.2015, PlanckXIII.2015}
we have for the first time a sufficient measurement of the polarization spectra
(both the temperature-E-mode polarization cross power spectrum (TE) and
the E-mode auto power spectrum (EE)) to clearly see multiple acoustic peaks
with well-defined locations. These locations provide a beautiful confirmation
of expectations for the response of the primordial plasma to small initial
adiabatic departures from complete homogeneity.

One can work out these expectations by solving the Einstein-Boltzmann equations
for evolution of the phase space distribution function of the various components
\citep[e.g.][]{Mukhanov1992, Ma1995, Seljak1996, Lewis2000,Lesg2011}. But these
numerical calculations, on their own, are not entirely satisfying. In addition
to knowing the answer, we want understanding.  This desire has led to many
papers aimed at an analytic understanding of the model power spectra
\citep[e.g.][]{Peebles1970, Doro1978, Atrio-Barandela1994, Seljak1994,
Jorgensen1995, Hu1995a, Hu1995b,  Hu1996, Hu1996a, Zaldarriaga1995, Hu1997a,
Weinberg2001, Weinberg2001a, Weinberg2002, Mukhanov2004, Bartolo2007, Cai2012}.
In this article, motivated by the recent first measurements of TE and EE extrema
locations, we develop a detailed analytic understanding of the locations of the
peaks in TT and EE, and the extrema in TE in the context of $\Lambda$CDM model.

The peak structure itself has drawn special attention. Back to about 15 years
ago, when only the first  TT peak was readily measured \citep[e.g.][]{Scott1994,
Smoot1997, Hancock1998,Lineweaver1998, Bond1998, Bond1999,Bond1999a,Miller1999,
Efstathiou1999, Tegmark1999, Tegmark2000, Tegmark2000a, Bond2000, Bond2000a,
Knox2000a,DeBernardis2000, Pierpaoli2000}, it was found to be consistent with
the  standard  $\Lambda$CDM model with adiabatic initial conditions, and imposed
tight constraints on other competing models  for example, $\Lambda$CDM model
with isocuravture initial conditions and topological defect models
\citep[e.g.][]{Hu1995b, Turok1996b,Turok1996a,Mag1996}. With more peaks measured
in recent years
\citep{Jaffe2001, de2002, Page2003, Benoit2003, Durrer2003, Readhead2004,
Jones2006, Hinshaw2007a, Cora2008, Preke2009, Naess2014}, the topological defect
models were ruled out \citep{Albrecht2000}, and the constraints on
the isocurvature modes have been improved to unprecedented precision
\citep[e.g.][]{Bucher2000,Trotta2001,Amendola2002,Bucher2002, Moodley2004,
Bean2006, Komatsu2009, Komatsu2011,Hinshaw2013a, PlanckXXII.2014}.

Part of the beauty of the TT, TE and EE measurements is that a very simple
analytic model provides us with a qualitative understanding
of the observed features. In the next section we will define this model and use
it to produce `baseline' predictions for the peak locations. It works especially
well for the relative locations of the peaks. For example, the temperature
anisotropies are predominantly sourced by temperature fluctuations at the last
scattering surface (LSS). We expect that the standing-wave modes that have hit
an extremum in temperature contrast right at the epoch of last scattering, will
be at a null in their peculiar velocities. Further, since gradients in peculiar
velocities are the dominant source of polarization anisotropy, peaks in TT
should correspond to minima in EE. This is roughly what we observe.

But the above picture is discrepant, in detail, with observations and with the
expectations of the $\Lambda$CDM model. To achieve an understanding that is
quantitatively correct, at a level consistent with the precision of current
measurements, we have to take into account a number of effects. We have found
that all these factors are important:  time-varying gravitational potentials
that are still non-zero at last scattering, neutrino free-streaming, the failure
of the tight-coupling approximation, the shape of the primordial power spectrum,
details of the projection from three dimensions to two, and the finite width of
the LSS. We work out, sometimes analytically, mostly by numerical methods, the
contribution of each one of these effects to the shifting of each of the peaks
from their locations in the baseline model.

The paper is organized as follows. In Section \ref{sec:baseline}, we introduce
a baseline model interpreting the evolution of photon perturbations and the
power spectra based on the tight coupling approximation and simplified projection.
In Section \ref{sec:tca}, we first analytically derive the phase shifts of photon
perturbations induced by decoupling, gravitational potential transient and
free-streaming neutrinos, then numerically test the analytic results by
examining the evolution of a single $k$ mode. In Section \ref{sec:atstar}, we
numerically measure the phase shift of the photon perturbations at the LSS,
single out the contribution from each effect, and analytically interpret them.
In Section \ref{sec:proj}, we investigate the impact of  projection on the peak
locations in details. We compare the peak locations determined from \Planck
measurements to expectations under the $\Lambda$CDM model in Section
\ref{sec:measure}, and  conclude in Section \ref{sec:conclusion}.

In this paper, we will work in the conformal Newtonian gauge
\be
ds^2 = a^2(\eta)\l[-(1+2\Phi) d\eta^2 + (1-2\Psi)\delta_{ij} dx^idx^j \r],
\ee
where $\eta$ is the conformal time, and the scalar perturbation $\Psi$ and $\Phi$
are related to the convention of \citet{Dodelson03} by
$\Phi = \Psi_{\rm Dodelson}, \Psi = -\Phi_{\rm Dodelson}$.
The fiducial cosmology used in the paper is
the  best fitting flat $\Lambda$CDM cosmology from
\emph{Planck} TT+low P+lensing \citep{PlanckXIII.2015}.

\section{Baseline Model}
\label{sec:baseline}
In this section, we will construct a simple analytic model, our baseline
model, that predicts the peak locations. This simple model neglects many
important effects. Much of the rest of the paper is then devoted to explaining
the differences between the approximate predictions
of this baseline model, and the numerically calculated, essentially exact predictions.

\subsection{Before Recombination}
The evolution of a photon-baryon plasma is governed by the Einstein-Boltzmann
equations, e.g., Eqs.(4.100 - 4.107) of \citet{Dodelson03},
\bea
\label{eq:EB1}
\Theta-\Theta_0-\mu V_b + \frac{\Pi}{2} P_2(\mu)  &=& \frac{ \dot \Theta - \dot\Psi + i k \mu (\Theta +\Phi)}{\dot \tau}, \\
\label{eq:EB2}
\Theta_p - \frac{\Pi}{2} \l( 1-P_2(\mu) \r) &=& \frac{ \dot \Theta_p + i k \mu \Theta_p }{\dot\tau}, \\
\label{eq:EB3}
V_b + 3 i \Theta_1 &=& \frac{R_b\l(\dot V_b + \frac{\dot a}{a} V_b + i k \Phi\r )}{\dot\tau},\
\eea
where $V_b$ is the bulk velocity of baryons, $\Theta_p$ is the strength of the
polarization field, $\Pi = \Theta_2 + \Theta_{p2} + \Theta_{p0}$,
$\tau(\eta)$ is the optical depth for a photon emitted at time $\eta$ and
received at today $\eta_0$, $R_b(= 3 \rho_b/ 4\rho_\gamma)$ is roughly the
ratio of baryon density over photon density, the dot denotes the derivative
with respect to the conformal time $\eta$, and $\mu = \hat k \cdot \hat p $ is
the cosine of the angle subtended by the wavevector $\vec k$ and the photon
propagation direction $\vec p$. To be clear, we adopte the most commonly
used convention of Legendre multipoles,
$\Theta (\mu) = \Sigma_{\ell=0}^{\infty} (-i)^\ell (2\ell+1) \Theta_\ell P_\ell(\mu)$
in Eq.~(\ref{eq:EB3}), and we neglecte the small corrections induced by
the nonzero sound speed of bayrons $c_b^2\sim T_b/\mu_b$, where
$T_b$ is the temperature of baryons and $\mu_b$ is the mean molecular weight
\citep[see e.g.][for details]{Ma1995}.

In the tight coupling limit, the first few multipoles can be obtained by
perturbative expansion with respect to $k/\dt$, which is expected to be much
smaller than unity before decoupling. Expanding Eqs. (\ref{eq:EB1}-\ref{eq:EB3})
 to $O(k/\dt)$, we get \citep[also see][]{Hu1995a, Hu1995b,  Hu1996,Hu1996a,
 Zaldarriaga1995, Seljak1996}
\bea
\Pi = \frac{5}{2} \Theta_2, \Theta_2 = -\frac{8}{15}\frac{k}{\dt} \Theta_1, \Theta_1 = \frac{i}{k}\l(\dot\Theta_0-\dot\Psi \r),
\eea
and the monopole satisfies
\be
\label{eq:monopole}
\l\{\frac{d^2}{d\eta^2} + k^2c_s^2\r\}[\Theta_0-\Psi] = -\frac{k^2}{3}\l(\Phi +\Psi \r),
\ee
where $c_s=1/\sqrt{3(1+R_b)}$ is the sound speed of the photon-baryon plasma
and we have dropped a small correction $\sim R_b$ in the above equation.
The monopole is actually a simple harmonic oscillator forced by gravitational driving.
Potentials $\Phi$ and $\Psi$ decay rapidly inside horizon during radiation domination,
and keep constant during matter domination.
For simplicity we drop, for now, the $\ddot\Psi$ term on the left side of the above equation
and we have \citep{Hu2002c}
\be
\l\{\frac{d^2}{d\eta^2} + k^2c_s^2\r\}[\Theta_0+\Phi] = 0,
\ee
where we have used the facts that $\Phi=\Psi$ in the absence of photon anisotropic stress,
$c_s^2 \simeq 1/3$, and $\ddot\Psi$ term is small after potentials decay.
Assuming adiabatic initial conditions, expected from the simplest inflationary models,
$[\dot\Theta+\dot\Phi](\eta=0)=0$, we obtain
\bea
\label{eq:harmo}
\TP \propto \cos(kr_s),\quad \Theta_1 \propto \sin(kr_s),\quad \Pi\propto\frac{k}{\dt}\sin(kr_s),
\eea
where $r_s(\eta) =\int_0^\eta c_s d\eta $ is the sound horizon at time $\eta$.

\subsection{After Recombination}
After recombination, photons freely stream. Hence the temperature anisotropies
we observed today are largely determined by the photon perturbation at the LSS,
$\Theta(\vec x = 0, \hat \gamma, \eta = \eta_0 )
\simeq \TP(\vec x = \hat\gamma(\eta_0-\eta_\star), \hat\gamma, \eta = \eta_\star)$,
\footnote{Strictly speaking, it is more appropriate to write
$[\Theta+\Phi](\vec x = 0, \hat \gamma, \eta = \eta_0 )
\simeq \TP(\vec x = \hat\gamma(\eta_0-\eta_\star), \hat\gamma, \eta = \eta_\star)$,
but the local potential today $\Phi(\vec x = 0, \eta=\eta_0)$ has no direction dependence,
thus has no influence on the CMB anisotropies. }
where $\hat\gamma$ is the observation direction, $\eta_0$ is the conformal time
today and $\eta_\star$ is the conformal time of the LSS. To study the statistical
property of the anisotropies, we usually expand the field in terms of spherical harmonics
\be
a_{\ell m} = \int d\Omega \ Y_{\ell m}(\hat \gamma) \ \Theta(\vec x = 0, \hat \gamma, \eta = \eta_0),
\ee
and define the temperature power spectrum $\Cl\equiv \left\langle a_{\ell m} \ a_{\ell m}^* \right\rangle.$
With some geometric transforms \citep[e.g.][]{Dodelson03}, the power spectrum is explicitly expressed as
\be
\Cl = \int dk \ k^2\ \Theta_\ell^2(k),
\ee
where $\Theta_\ell(k)$ is the multipole moment of the temperature field of $\vec k$ mode,
\be
\Theta_\ell(k) = \frac{1}{(-i)^\ell}\int_{-1}^{1} \frac{d\mu}{2} P_\ell(\mu)\  \Theta(\vec k, \hat\gamma, \eta_0),
\ee
where $\mu=\hat k \cdot\hat\gamma$.

Using the plane-wave pattern (see Fig. \ref{fig:proj})
\bea
\Theta(\vec k, \hat\gamma, \eta_0)
& \simeq &\TP(\vec k, \hat\gamma, \eta_\star) \nn\\
& =& \TP(k,\eta_\star)\times e^{i  \hat \gamma \cdot \vec k  (\eta_0-\eta_\star)},
\eea
where $\TP(k,\eta_\star)$ is the oscillation amplitude [Eq.~(\ref{eq:harmo})]
and $e^{i  \hat \gamma \cdot \vec k  (\eta_0-\eta_\star)}$ is the spatial pattern,
we have
\bea
\label{eq:bessel}
\Theta_\ell(k)
&\simeq& \frac{1}{(-i)^\ell}\int_{-1}^{1} \frac{d\mu}{2} P_\ell(\mu)\  \TP(k,\eta_\star)e^{i \mu k (\eta_0-\eta_\star)}, \nn\\
&=& (-1)^\ell \TP(k,\eta_\star) j_\ell[k(\eta_0-\eta_\star)],
\eea
where $j_\ell[k(\eta_0-\eta_\star)]$ is a spherical Bessel function, which peaks
at $\ell\simeq k(\eta_0-\eta_\star)$. Therefore, we expect the TT power at mode
$\ell$ is mainly sourced by $\TP(k,\eta_\star)$ by mode
$k\simeq \ell/(\eta_0-\eta_\star)$.
Similar argument yields that EE and TE are mainly sourced by $\Pi(k,\eta_\star)$
and $[(\Theta_0+\Phi)\times\Pi](k,\eta_\star)$, respectively \citep[see e.g.][]{Hu1995b,Zaldarriaga1995,Hu1997c,Tram2013}.
As a result,
\bea
\Dl &\sim&  \TP^2(kr_{s,\star})  \propto \cos^2(\ell \theta_\star), \nn\\
\Dle&\sim&  \Pi^2(kr_{s,\star})   \qquad\qquad \propto\sin^2(\ell \theta_\star), \nn\\
\Dlc&\sim&  \l\{\TP\times \Pi\r\} (kr_{s,\star})  \propto \sin(2\ell \theta_\star),
\eea
where $D_\ell^{\rm XX} \equiv \ell(\ell+1)/(2\pi) C_\ell^{\rm XX}$, with ${\rm XX= TT, TE, EE}$,
and $\theta_\star$ is the angular size of the sound horizon at recombination,
$\theta_\star \equiv r_{s,\star}/(\eta_0-\eta_\star) = 1.04\times10^{-2}$ \citep[e.g.][]{PlanckXVI.2013, PlanckXIII.2015}.
Therefore  $\Dl,\Dle, \Dlc$ reach their $p$-th peak at
\bea
\ell_p^0({\rm TT}) &=& 302\  p, \nn\\
\ell_p^0({\rm EE}) &=& 302 (p-0.5), \nn\\
\ell_p^0({\rm TE}) &=& 151 (p+0.5),
\eea
respectively (Throughout this paper, we refer to both the \emph{maxima} and the \emph{minima} in the TE power spectrum as \emph{peaks} due to  the arbitrary sign of E mode, and we also refer to the zero points of the TE power spectrum as \emph{troughs}). We are only interested in the peaks in the power spectra which correspond to the extrema of sources $\TP$, $\Pi$, $\TP\times\Pi$, and so carry phase information of the acoustic oscillation. The troughs in the spectra corresponding to the zero points of the sources also carry phase information, but baryon drag shifts the zero points and introduces extra uncertainty. Therefore the troughs in the spectra, and the reionization bumps in EE and TE power spectra are not investigated in this paper.

\begin{figure}
\includegraphics[scale=0.44]{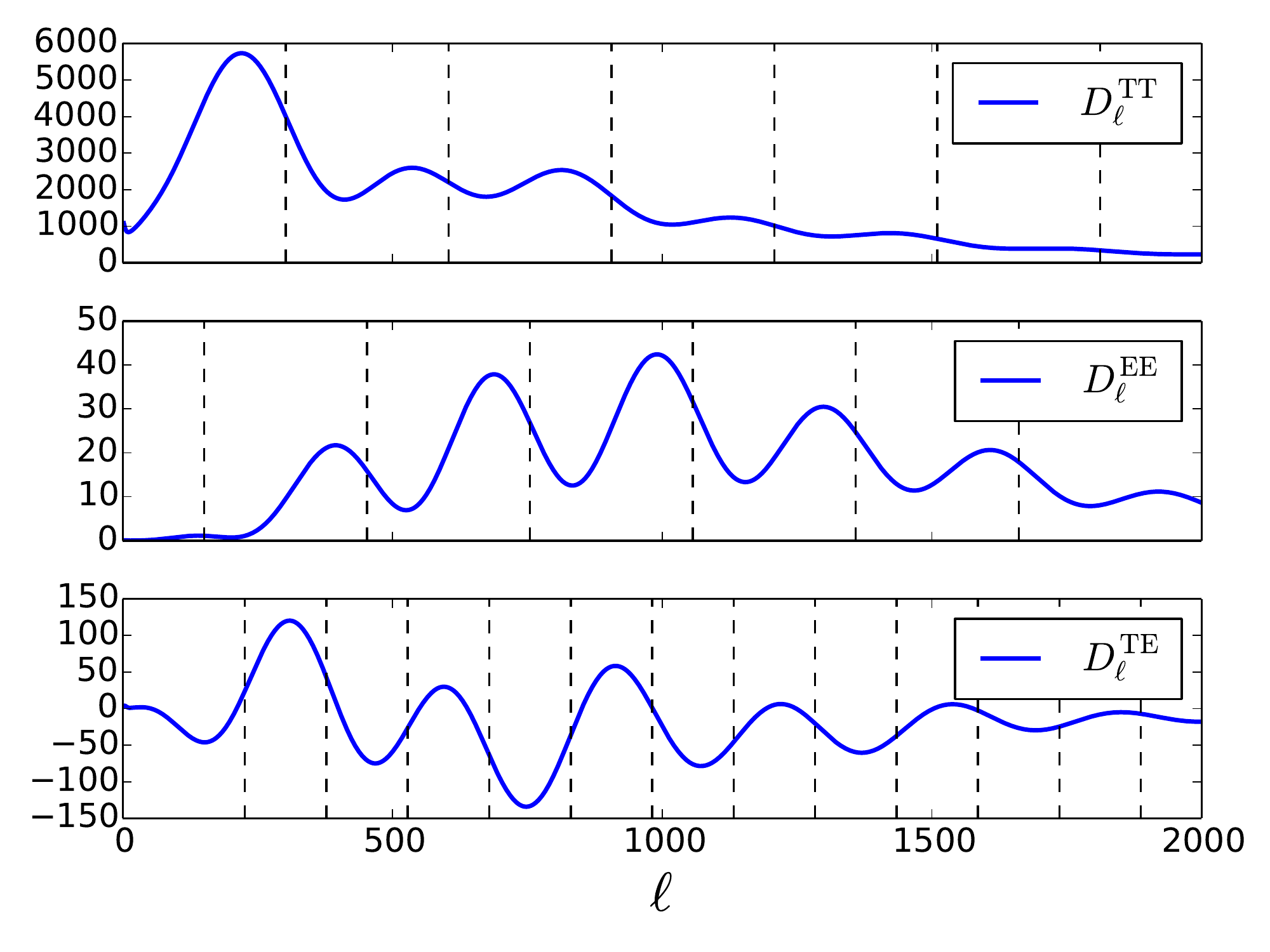}
\caption{\label{fig:baseline} Comparison of the spectra of the fiducial cosmology (solid curves) and the peak locations predicted by the baseline model (vertical dashed lines). }
\end{figure}

In Fig. \ref{fig:baseline}, we compare the theoretical spectra of the fiducial cosmology with the baseline model. The baseline model is roughly correct in its prediction for peak spacing and for the relative locations of the peaks in different spectra. But  peak locations predicted by the baseline model  do not coincide with the true locations, and the typical phase shift $\delta\ell_p$ is about one fourth of the oscillation period. In addition, the baseline model also predicts that EE peaks are located halfway between TT peaks and also halfway between TE peaks, which is not exactly true either.

Despite its deficiencies, we find the baseline model to be a useful starting point.
In the remainder of the paper we explain the differences between these baseline predictions
and the predictions of the $\Lambda$CDM  model when calculated much more precisely.

\section{Evolution of phase shifts in the photon perturbations}
\label{sec:tca}

According to the baseline model, $\Theta_0 (\Theta_1)$ can be described as a
simple harmonic oscillator under two assumptions: tight coupling between photons
and baryons, and negligible impact of gravitational driving. In fact, both the
decoupling effect and decaying gravitational potentials affect the amplitude and
the phase of the acoustic oscillation. Taking these into account,
we may formally write the solution as
\be
\label{eq:3phi}
\TP  \propto \cos(kr_s + \phi_{\rm tot}),
\ee
where $\phi_{\rm tot} \equiv  \phi_{\rm dcp} + \phi_{\rm gr} $ with $\phi_{\rm dcp},
\phi_{\rm gr}$ denoting the phase shift induced by decoupling and gravitational
driving, respectively. The latter can be further decomposed as
$\phi_{\rm gr}= \phig +  \phin$, due to the fact that the decay of $\Phi+\Psi$
is caused by photon pressure and neutrino free-streaming.  To distinguish them,
we call $\phig$ as gravitational potential transient induced phase shift, and
call $\phin$ as neutrino induced phase shift. The reason for this decomposition
shall be clear later.

In the remainder of this section, we will analytically derive the phase shift
induced by each effect and numerically measure these phase shifts.

\subsection{Decoupling: $\phi_{\rm dcp}$}

After a mode enters the horizon ($k\eta \gtrsim 1$), the tight coupling
approximation becomes less reliable, and the small decoupling effect induces
both diffusion damping and phase shift to the evolution of photon perturbations.
The diffusion damping was analytically studied in
\citep[e.g.][]{Silk1968,Hu1995a,Zaldarriaga1995} by expanding the correction to
tight coupling approximation to $O(k/\dt)^2$. For our purpose of exploring the
phase shift induced by decoupling, we extend the correction to $O(k/\dt)^3$ and
find the analytic expression of the phase shift induced by decoupling
$\phi_{\rm dcp}$ (see Appendix \ref{app:dcp}).
Consequently, the intervals $\Delta(kr_s(\eta_p))\equiv kr_s(\eta_{p})-kr_s(\eta_{p-1})$
of the $p$-th and $(p-1)$-th extrema in $\Theta_0(k,\eta)$
are no longer equal to $\pi$. Instead, $\Delta(kr_s) = \pi - \Delta(\phi_{\rm dcp})$.
The intervals $\Delta(kr_s)$ measured from the  Boltzmann code {\tt Class} \citep{Lesg2011, Blas2011}
and obtained from the analytical result in Appendix \ref{app:dcp} are shown in Fig. \ref{fig:intervals}.
They are in good agreement except at late time when the correction to tight
coupling up to $O(k/\dt)^3$ is no longer accurate and at early time when the gravitational driving is important.

\begin{figure}
\includegraphics[scale=0.44]{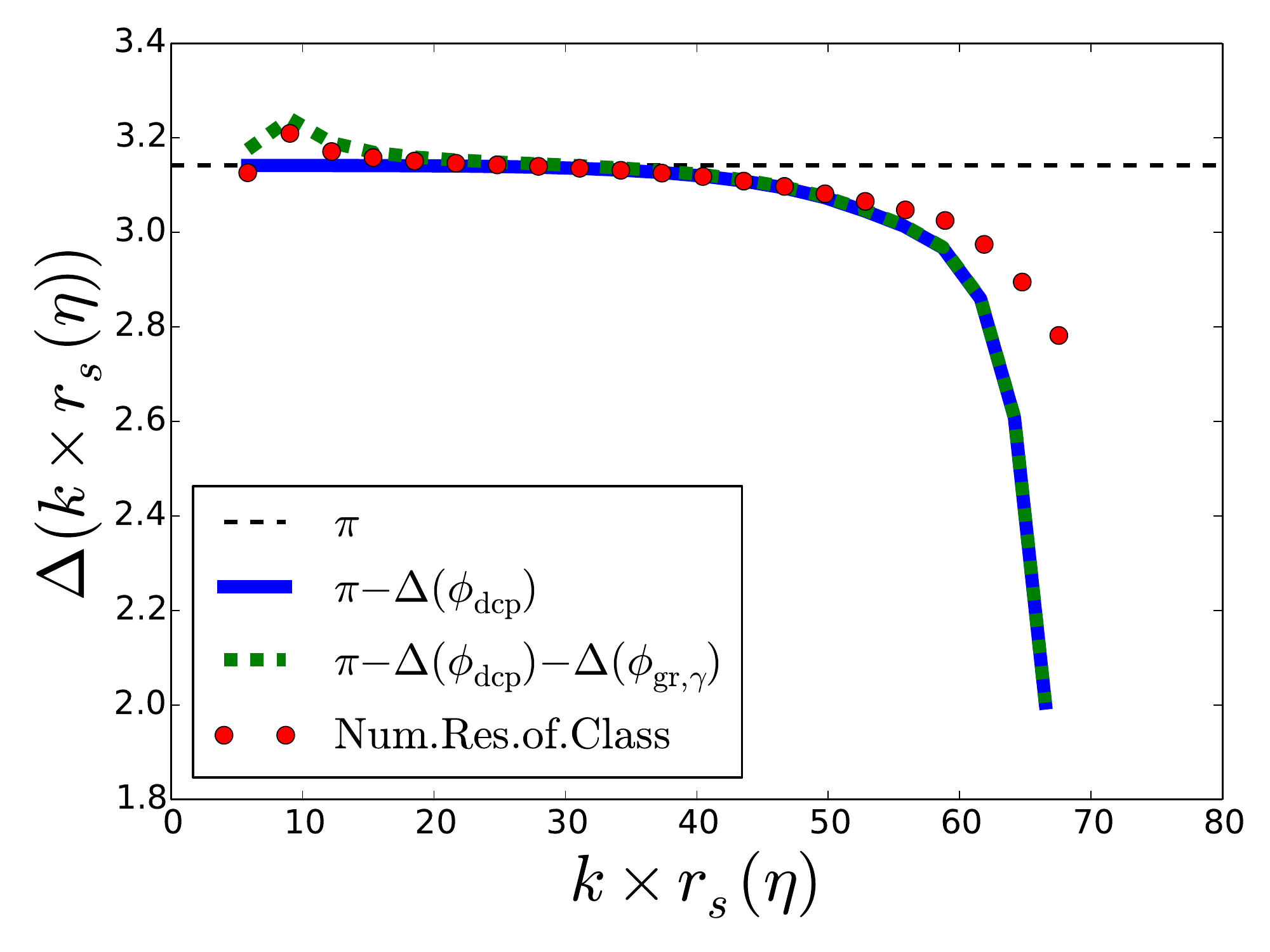}
\caption{\label{fig:intervals} The intervals $\Delta(kr_s)$  of neighboring
peak-trough of $\l[\Theta_0+\Phi\r](k, \eta)$ for mode $k=0.5 \ {\rm Mpc}^{-1}$.
Dots are numerical results of peak-trough intervals.
Solid line is the analytic result of high-order correction to the tight coupling approximation $\phi_{\rm dcp}$.
Dashed line is the result of corrections from both late-time high-order correction $\phi_{\rm dcp}$
and early-time gravitational driving $\phi_{\rm gr,\gamma}$ sourced by photon perturbations. }
\end{figure}

\subsection{Transient: $\phig$}
Both the amplitude and the phase of the acoustic oscillation are modulated by
the gravitational driving at early time when the potentials have not decayed.
Following \citet{Bashinsky2004}, we define the overdensity of photon number with
respect to coordinate volume,
$\dg \equiv 3(\Theta_0 - \Psi$) and  two potentials $\Phi_\pm = \Phi \pm \Psi$,
which satisfy the dynamical equation accurate  to $O(k/\dt)$ [Eq. (\ref{eq:monopole})],
\be
\label{eq:driving}
\dg'' +\dg = -3\Phi_+,
\ee
where the prime denotes differentiation with respect to $kr_s$.
Assuming negligible neutrinos,
both $\Phi_\pm$ and $\dg$ can be analytically solved during radiation domination,
$\dg \propto  \cos(kr_s + \theta(kr_s))$,
where the phase shift $\theta(kr_s)$ decays with time as
\be
\label{eq:transient}
\theta(kr_s) \simeq \frac{2}{kr_s}\Big|_{kr_s\gg1},
\ee
(see Appendix \ref{app:phig} for accurate expressions).
We have assumed radiation domination in above derivation, so we expect it is
correct only for $\eta\lesssim\eta_{\rm eq}$. After the transition to matter domination,
the gravitational potentials keep roughly constant, and $\theta$  also freezes at
\be
\label{eq:freeze}
\theta(kr_s)\Big|_{\eta > \eta_{\rm eq}} \simeq \quad \theta(kr_s)\Big|_{\eta = \eta_{\rm eq}}.
\ee

The above analysis shows that, for large $k$, $\theta(kr_s)$ decays with time
to zero, as the potentials decay to zero; while for small $k$, $\theta(kr_s)$
does not decay to zero, as the potentials do not completely decay. To summarize,
$\theta(kr_s)$ traces the potential transient. This also explains why we call
$\theta(kr_s)$ as the gravitational potential transient induced phase shift.

A minor point is that $\theta$ is the phase shift for $\dg = 3(\Theta_0 - \Psi)$,
while the quantity more relevant to the TT power spectrum  is the effective
temperature perturbation $\TP$. Hence what we plot in Fig. \ref{fig:intervals}
is the peak-trough spacing of $\TP$, i.e., $\phig$ instead of $\theta$.
According to Fig. \ref{fig:intervals}, the transient induced phase shift $\phig$
accounts for most of the residual phase shifts of the first few extrema in $\TP$.

\subsection{Neutrinos: $\phin$}
Another important component during radiation domination is free-streaming neutrinos,
which recast the potential transient and  introduces a new phase shift component $\phin$.
Under assumption of radiation domination (neglecting matter and dark energy),
potential $\Phi_+(kr_s\rightarrow\infty)$ completely decays, and so does the
transient induced phase shift $\phig(kr_s\rightarrow\infty)$.
More generally, previous studies \citep{Bashinsky2004, Bashinsky2007, Baumann2016}
show that a nonzero phase shift $\phi_{\rm gr}(kr_s\rightarrow\infty)$ requires
modes propagating faster than the sound speed of photon-baryon plasma $c_s$.
Neutrinos freely stream in the light speed, $c > c_s$, so a nonzero phase shift
$\phin(kr_s\rightarrow\infty)$ is expected. Accurate to $O(R_\nu)$,
\citet{Bashinsky2004} and \citet{Baumann2016} obtained a scale-independent phase shift
\be
\phin(kr_s\rightarrow\infty) = 0.191 R_\nu \pi,
\ee
where $R_\nu = \rho_\nu/(\rho_\nu+\rho_\gamma)$ is the energy fraction of
neutrinos in the radiation. Taking account of matter domination, the above
scale-independent result only applies for modes entering the horizon during
radiation domination; while for modes entering the horizon during matter
domination, $\phin$ approaches zero as $\sim k^2$ (see Appendix \ref{app:phin} for details).

In contrast to $\phig$, neutrino induced phase shift $\phin$ does not the trace
potential transient, though neutrinos indeed affect the transient.

\section{Phase shifts in photon perturbations at the LSS}
\label{sec:atstar}
In this section, we numerically measure the phase shift of photon perturbations
at the LSS, single out the contribution from each effect, and analytically interpret them.

\begin{figure}
\includegraphics[scale=0.42]{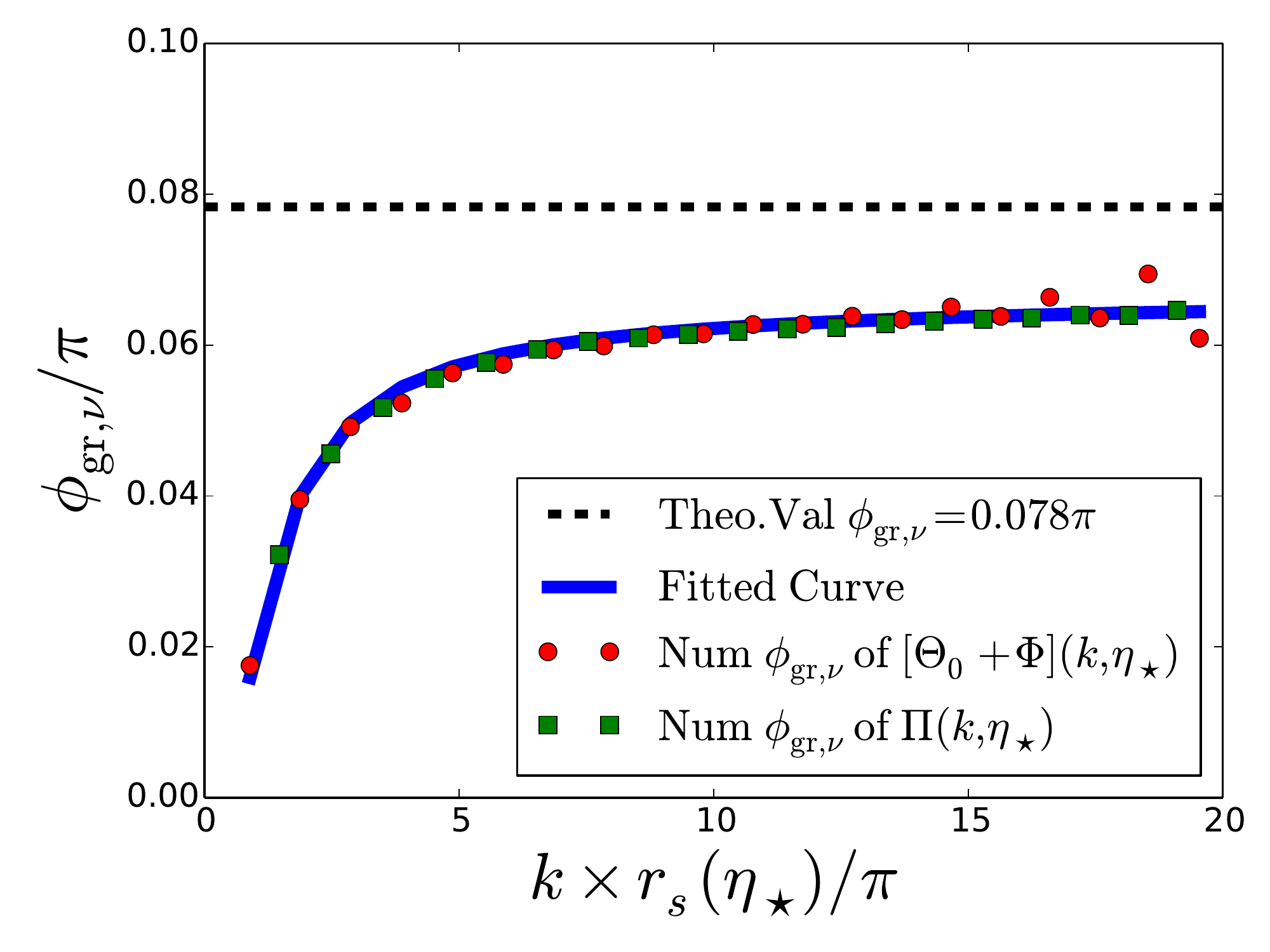}
\caption{\label{fig:neu} Phase shifts of sources induced by $3.046$ neutrinos and measured at the LSS.}
\end{figure}

\begin{figure*}
\includegraphics[scale=0.4]{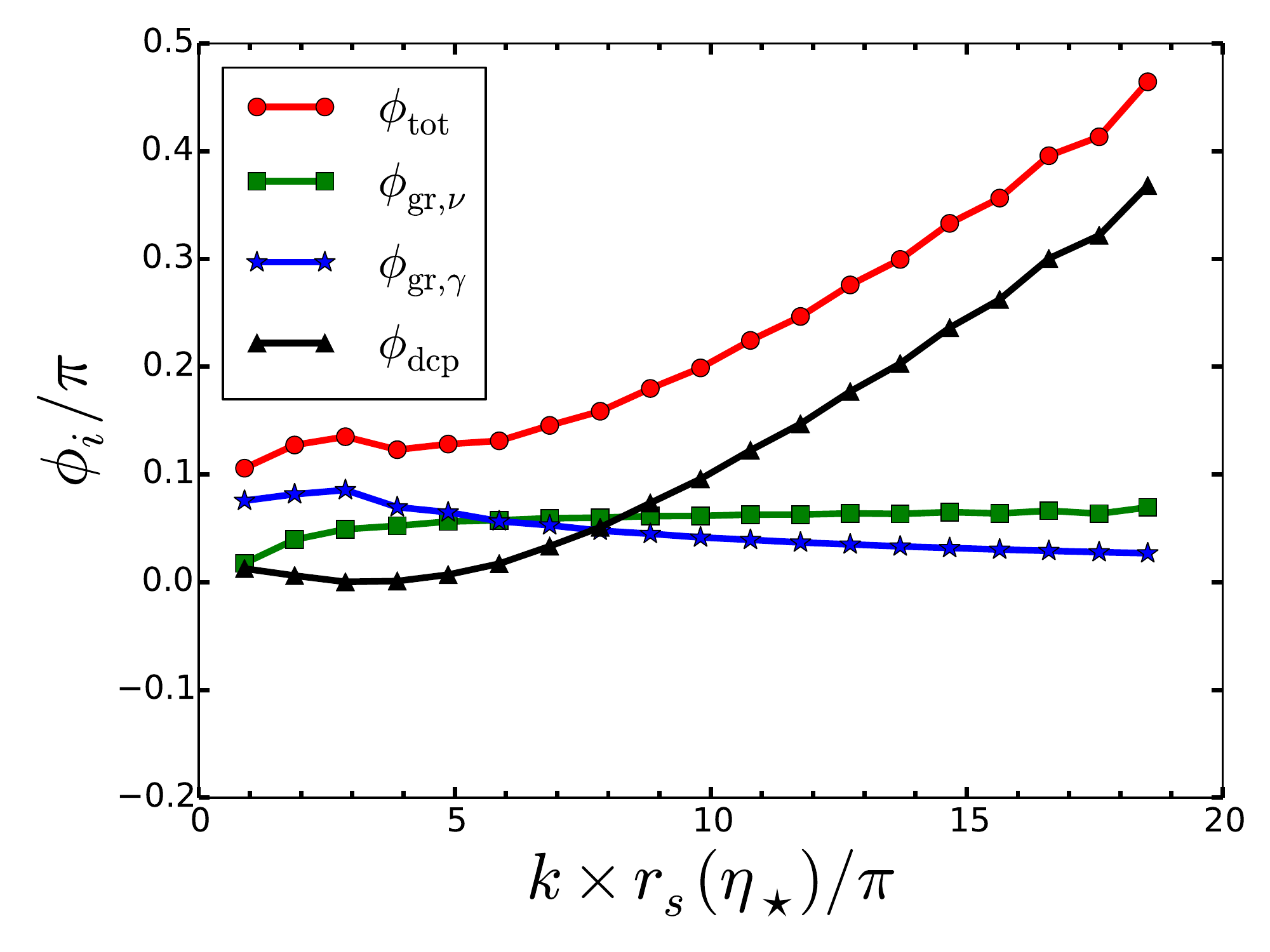}
\includegraphics[scale=0.4]{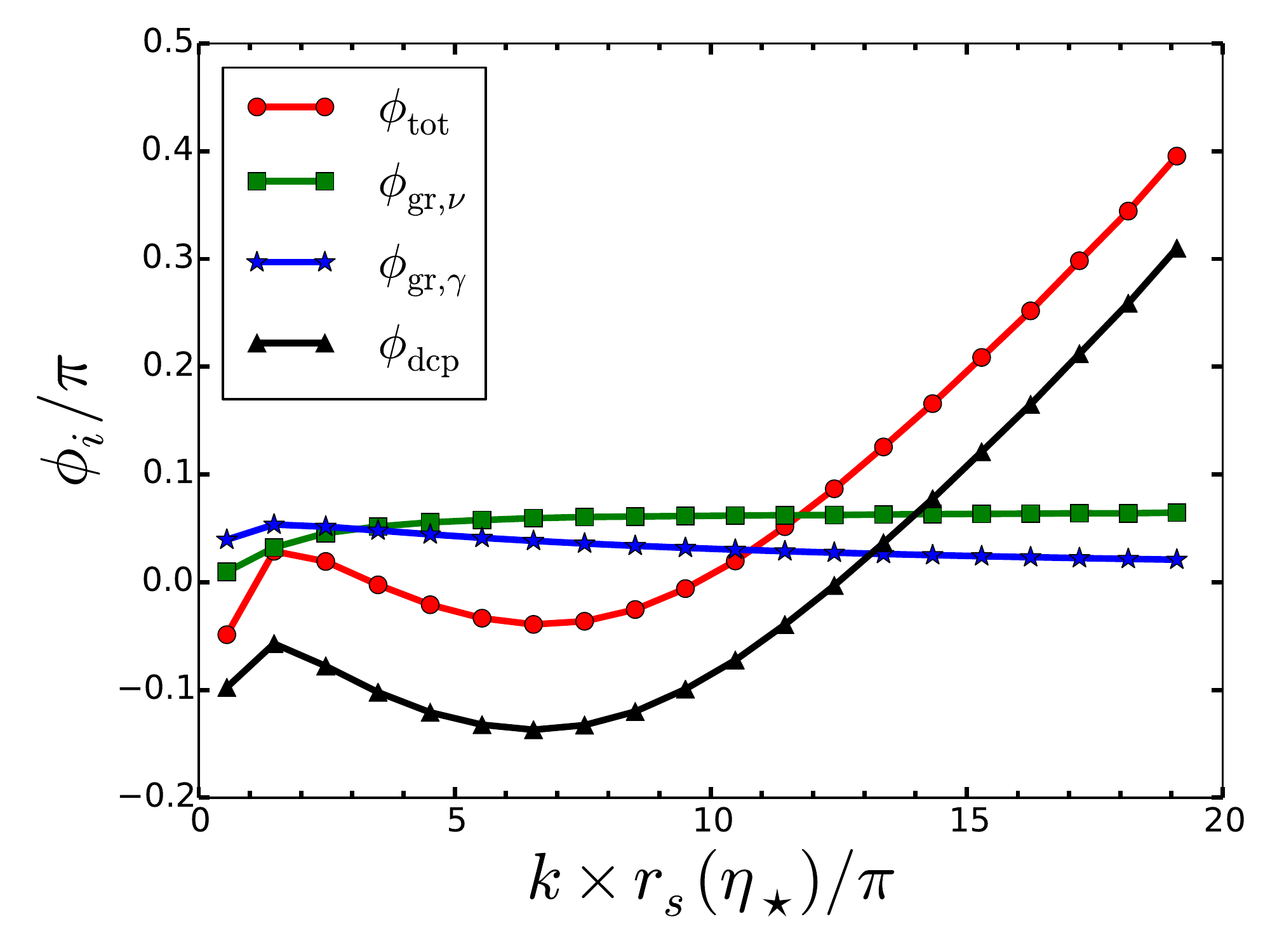}
\caption{\label{fig:atstar} The phase shifts of $\TP$ (left panel) and of $\Pi$
(right panel) induced by different physical effects measured at the LSS.}
\end{figure*}

To measure the total phase shift $\phi_{\rm tot}$ of the monopole source
$\TP(kr_{s,\star})$ for the fiducial cosmology, we fix $\eta=\eta_\star$ and
match its extrema as a function of $k$ to those of $\cos(kr_{s,\star} +\phi_{\rm tot})$
by adjusting $\phi_{\rm tot}$. In a similar way, the total phase shift
$\phi_{\rm tot}[\Pi]$ of the polarization source $\Pi(kr_{s,\star})$ is also
measured by matching with $\sin(kr_{s,\star} +\phi_{\rm tot}[\Pi])$.

To single out the neutrino induced phase shift $\phin$ from the total phase shift,
we need to filter out the other two effects. For this purpose, we construct a
comparison cosmology without neutrinos and with
$z_{\rm eq}, \theta_\star,\theta_D,\omega_{\rm b}$ fixed to the corresponding
values of the fiducial cosmology, by adjusting the cold dark matter density
$\omega_{\rm c}$, dark energy fraction $\Omega_\Lambda$ and the helium fraction
$Y_{\rm P}$  \citep{Follin2015}. Then we measure the monopole source
$\TP(kr_{s,\star})$ at the LSS for both the fiducial cosmology and
the comparison cosmology. The displacement between the extrema locations of
the two is $\phin$, which is plotted in Fig. \ref{fig:neu}. As expected, $\phin$
approaches  zero for small $k$ modes and approaches  a constant for large $k$ modes.
We find  $\phin(k\rightarrow \infty) = 0.067\pi$, which is about $15\%$ lower
than the lowest-order analytically derived value $0.078\pi$. Note that the
$\phin$ derived in Appendix \ref{app:phin} is actually the phase shift of the
monopole $\phin[\Theta_0]$, not the phase shift of the polarization $\phin[\Pi]$.
Both of them (and also $\phin[\Theta_1]$) are well fitted by
\be
\phin(k ,\eta_\star) = \frac{1}{7.5} \tan^{-1}\l(\frac{kr_{s,\star}}{\pi} -0.5\r),
\ee
at least for $kr_{s,\star} \gtrsim \pi$.

To single out the gravitational potential transient induced phase shift $\phig$,
we solve the Einstein-Boltzmann equations in the strict tight coupling limit
($\Theta_{\ell\geq2} = 0$) for the comparison cosmology constructed above, and
evaluate the monopole at $\eta_\star$. In this way, we get rid of both
$\phi_{\rm dcp}$ and $\phin$, therefore the only phase shift left is $\phig$
(see left panel of Fig. \ref{fig:atstar}).
Analytic study [Eqs.(\ref{eq:transient},\ref{eq:freeze})] shows that
\be
\phig(k,\eta_\star) \simeq \frac{2}{kr_s(\eta_{\rm eq})}\Big|_{kr_s(\eta_{\rm eq})\gg1},
\ee
which is consistent with the numerical result for large $k$ modes
(left panel of Fig. \ref{fig:atstar}). The transient induced phase shift
$\phig[\Pi]$ in the polarization source is more subtle to tap, because the
polarization source $\Pi\simeq - (4k/3\dt) \Theta_1$ vanishes in the strict
tight coupling limit. We tentatively extract $\phig[\Pi]$ by matching the
extrema in $k\times \Theta_1(kr_{s,\star})$ with those of
$\sin(kr_{s,\star} + \phig[\Pi])$ (see right panel of Fig. \ref{fig:atstar} for the numerical results).

With $\phin$ and $\phig$ singled out, the residual part is certainly the
decoupling induced phase shift $\phi_{\rm dcp}$ given by $\phi_{\rm tot}-(\phin+\phig)$,
which as expected scales as $\sim k^3$ for small $k$ modes
(see left panel of Fig.\ref{fig:atstar}).  The decoupling induced phase shift
$\phi_{\rm dcp}[\Pi]$ in the polarization source is also extracted in the same
way, which shows more structures than that of the monopole (see right panel of
Fig.\ref{fig:atstar}). According to the analytic study in Appendix
\ref{app:dcp}, $\phi_{\rm dcp}[\Pi]$ scales as $\sim O(k^3)-O(k)$, where the
former term is the same to $\phi_{\rm dcp}$ of the monopole and the latter
term comes from the fact that $\Pi$ and $\Theta_0$ are out of phase by slightly
less than 90 degrees. The scaling explains the overall shape of
$\phi_{\rm dcp}[\Pi]$. The `anomaly' of the first point is due to the rise in
the amplitude of $\Pi$ as $k$ increases. The polarization is sourced by the
gradient of the velocity field, $\Pi\propto k \Theta_1$, and so the factor of
$k$ drives the extrema in $\Pi$ to larger $k$ modes. It is straightforward to
estimate that the first extremum is driven away by $\delta(kr_s)\simeq 0.1\pi$,
which is exactly the anomaly dip we observed, and the effect on other extrema
is weak as their changes in $k$ have a smaller dynamic range.

\section{Projection}
\label{sec:proj}
With our description of the phase shift $\phi_{\rm tot}$ in sources at the LSS complete,
we are ready to study the peak shift in the spectra.
In this section,  we first give a more rigorous treatment of the projection process
from photon perturbations at the LSS to the power spectra, then point out the
corrections to the baseline model, and  measure the peak shift induced by each correction.

\subsection{A Rigorous Treatment of Projection}
In the baseline model, the pictorial argument of projection yields a qualitative
understanding on the peak structure in the spectra. But for a quatitative understanding,
we need a rigorous treament of the projection process.
Let us start from the well-known line-of-sight solutions to Eq.~(\ref{eq:EB1})
\citep[e.g.][]{Hu1997c, Dodelson03},
\be
\label{eq:los}
\Theta(k,\mu,\eta_0) = \int_0^{\eta_0} d\eta \ \tilde S(k,\mu, \eta) e^{ik\mu(\eta-\eta_0)-\tau(\eta)},
\ee
where the source
\be
\tilde S(k,\mu, \eta) = \dot\Psi - ik\mu\Phi - \dot\tau \left[\Theta_0+\mu V_b -\frac{1}{2} P_2(\mu)\Pi \right].
\ee
From solution Eq.~(\ref{eq:los}), we obtain the multipoles
\bea
\label{eq:trsf}
\Theta_\ell(k)
&& =\int_0^{\eta_0} d\eta \ g(\eta) [\Theta_0(k,\eta)+\Phi(k,\eta)] j_\ell[k(\eta_0-\eta)] \nn\\
&& -\int_0^{\eta_0} d\eta \ g(\eta) \frac{iV_b}{k} \frac{d}{d\eta} j_\ell[k(\eta_0-\eta)] \nn\\
&& +\int_0^{\eta_0} d\eta \ g(\eta) \frac{3}{4}\frac{\Pi}{k^2}  \frac{d^2}{d\eta^2} j_\ell[k(\eta_0-\eta)]\nn\\
&& +\int_0^{\eta_0} d\eta \ e^{-\tau}  [\dot\Phi +\dot\Psi ] j_\ell[k(\eta_0-\eta)],
\eea
where $g(\eta)\equiv -\dot\tau e^{-\tau}$ is the visibility function which
narrowly peaks at the LSS $\eta=\eta_\star$, $\Theta_0(k,\eta)$ is the
amplitude we have investigated in Section \ref{sec:atstar} in detail.
The above rigorous formula not only verifies the naive expectation of the baseline model,
but also takes account of contributions from doppler effect (dipole),
polarization, and integrated Sachs-Wolfe (ISW) effect.
In addition, $j_\ell[k(\eta_0-\eta)]$ is nearly zero for
$k(\eta_0-\eta)\lesssim \ell$. The asymmetric projection can be understood
with the following pictorial argument.

\begin{figure}
\includegraphics[scale=0.5]{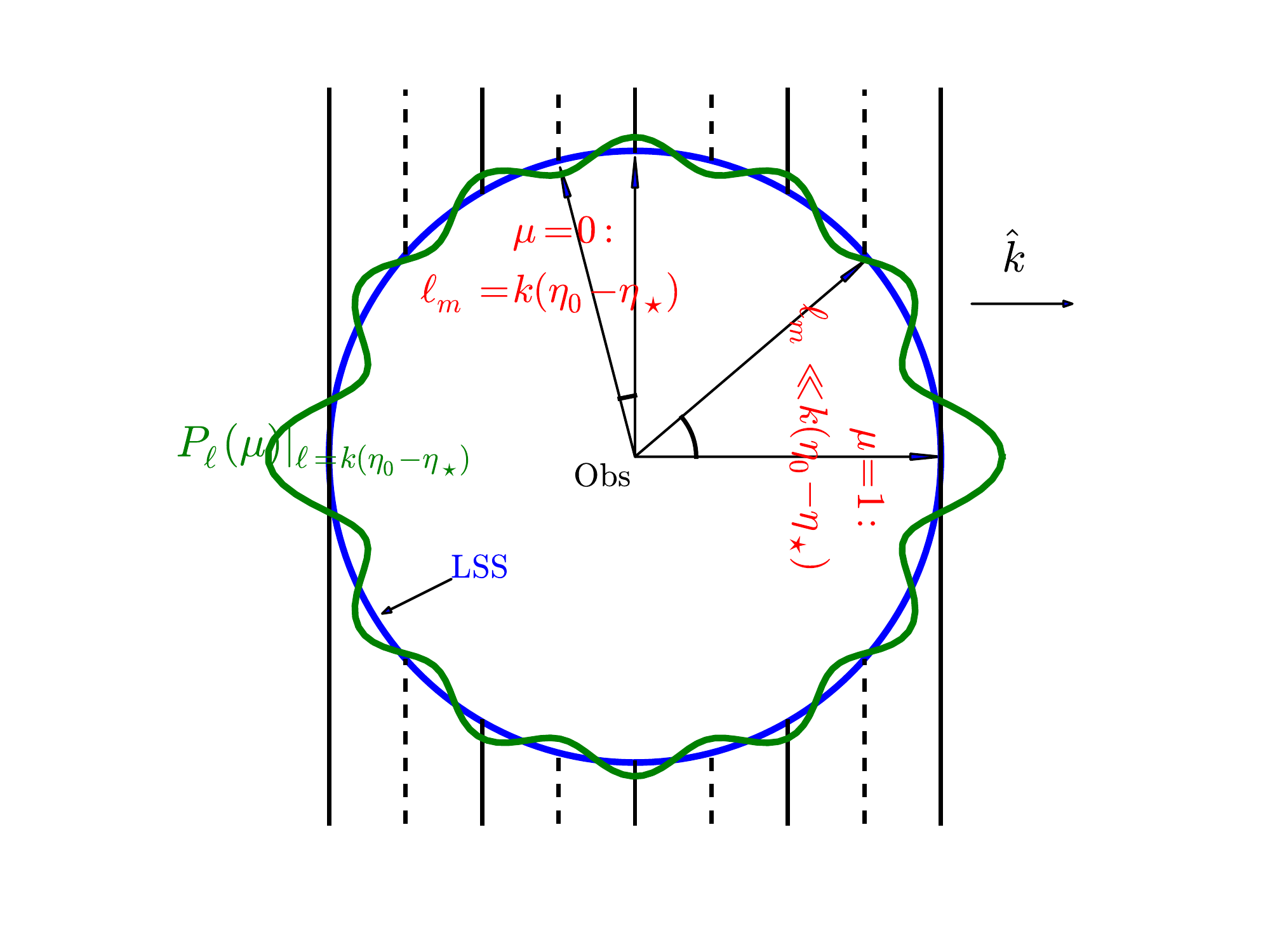}
\caption{\label{fig:proj} Illustration of the projection from three to
two dimensions. The round circle is the LSS, the vertical solid(dashed) lines
are the peaks(troughs) of mode $\vec k$ at $\eta_\star$. The wiggling curve
around the LSS is a Legendre polynomial $P_\ell(\mu)$ with $\ell=k(\eta_0-\eta_\star)$,
where $\mu = \hat k \cdot \hat \gamma$, is the cosine of the angle subtend by
the wavevector $\vec k$ and the direction of observation $\hat \gamma$.
In the $\mu=0$ direction, the peak-trough separation of the $\vec k$ mode
matches that of $P_{\ell_m}(\mu)|_{\ell_m =k(\eta_0-\eta_\star)}$ (shown).
In the $\mu=1$ direction, the peak-trough separation of the $\vec k$ mode
matches that of $P_{\ell_m}(\mu)|_{\ell_m \ll k(\eta_0-\eta_\star)}$ (not shown).}
\end{figure}

As shown in Fig.~\ref{fig:proj}, in the direction perpendicular to the wavevector, $\mu = 0$,
the peak-trough separation of the $k$ mode matches that of $P_{\ell_m}(\mu)$
with $\ell_m = k(\eta_0-\eta_\star)$.
Whereas in the direction parallel to the wavevector, $\mu = 1$, the peak-trough
separation subtends a larger angle and so is better matched with
a $P_{\ell_m}(\mu)$  with $\ell_m\ll k(\eta_0-\eta_\star)$. Therefore mode $k$
distributes its power on all modes satisfying
$\ell \lesssim k(\eta_0-\eta_\star)$. In other words, the power of mode $\ell$
is contributed by all modes satisfying $k (\eta_0-\eta_\star) \gtrsim \ell$ \citep[e.g.][]{Hu1997c}.

Using the transfer function $\Delta_\ell(k)$ defined by $\Delta_\ell(k)\equiv \Theta_\ell(k)/\Phi(k,0)$
and the primordial potential power spectrum $P(k)$ defined by
\be
\left\langle \Phi(\vec k, 0) \Phi^\star(\vec k', 0) \right\rangle = (2\pi)^3 \delta^{(3)}(\vec k -\vec k') P(k),
\ee
$\Cl$ is written as
\bea
\label{eq:tps}
\Cl  = \int dk \ k^2 P(k)\ \Delta_\ell^2(k),
\eea
up to some constant factor \citep[e.g.][]{Seljak1996, Dodelson03}.

\subsection{Corrections to the Baseline Model}

\begin{table*}
\caption{\label{table:tt} The shift of the $p$-th peak in the temperature power spectrum $\Dl$ defined by $\delta\ell_p \equiv 302\  p -\ell_p({\rm TT})$, consists of  $\delta\ell_{\rm monopole}$, $\delta\ell_{\rm non-monopole }$ and $\delta\ell_{\rm lensing}$, where the former can be decomposed as $\delta\ell_{\rm monopole} = \delta\ell[\phin]+ \delta\ell[\phig]+\delta\ell[\phi_{\rm dcp}]+\delta\ell[ k^2P(k)]+\delta\ell[j_\ell]+\delta\ell[g(\eta)]$.}
\begin{tabular}{c | c c c c c  c | c  c  c | l}
\hline \hline

{\rm $p$-th  peak} & $\delta\ell[\phin]$  & $\delta\ell[\phig]$  &  $\delta\ell[\phi_{\rm dcp}]$  &
$ \delta\ell[ k^2P(k)]$ & $\delta\ell[j_\ell]$ &
$\delta\ell[g(\eta)]$ & $\delta\ell_{\rm monopole}$ &
$\delta\ell_{\rm non-monopole }$ & $\delta\ell_{\rm lensing}$ & $\qquad\quad \delta\ell_p$ \\

\hline
{\rm 1st} & \phantom{0}5 & 23& 4 &  20\phantom{0} & 19 & $\phantom{-}7$ &   78  & $4$   	   & 0  & $82 = \phantom{1}302 - \phantom{1}220$\\
{\rm 2nd} & 12		 & 25& 1 &  7 		  & 29 & $\phantom{-}9$ &   83  & $-16$ 	   & 1  & $68 = \phantom{1}604 - \phantom{1}536$\\
{\rm 3rd} & 15		 & 26& 0 &  6 		  & 43 & $\phantom{-}7$ &   97  & $-6\phantom{0}$  & 2  & $93 = \phantom{1}906 - \phantom{1}813$ \\
{\rm 4th} & 16		 & 21& 0 &  4 		  & 40 & $\phantom{-}3$ &   84  & $-6\phantom{0}$  & 3  & $81 = 1208 - 1127$\\
{\rm 5th} & 17		 & 20& 2 &  3 		  & 45 & $-1$ 		&   86  & $-3\phantom{0}$  & 6  & $89 = 1510 - 1421$ \\
{\rm 6th} & 17		 & 17& 6 &  3 		  & 39 & $-9$ 		&   73  & $-1\phantom{0}$  & 15\phantom{0} & $87 = 1812 - 1725$\\
%{\rm 7th} & 18& 16& 10 &  2 & 41 & -16 & 71 &  4  & 33 & 108 = 2114 - 2006\\
\hline
\end{tabular}
\end{table*}

\begin{table*}
\caption{\label{table:ee} The shift of the $p$-th peak in  the polarization power spectrum $\Dle$ defined by $\delta\ell_p \equiv 302 (p-0.5) -\ell_p({\rm EE})$, consists of $\delta\ell_\Pi$ and $\delta\ell_{\rm lensing}$, where the former is decomposed as $\delta\ell_\Pi = \delta\ell[\phin]+ \delta\ell[\phig]+\delta\ell[\phi_{\rm dcp}]+\delta\ell[ k^2P(k)]+\delta\ell[j_\ell]+\delta\ell[g(\eta)]$. }
\begin{tabular}{c | c c c  c  c  c | c c | l }
\hline \hline
{\rm $p$-th peak} & $\delta\ell[\phin]$ & $\delta\ell[\phig]$  & $\delta\ell[\phi_{\rm dcp}]$  & $ \delta\ell[ k^2P(k)]$ & $\delta\ell[j_\ell]$ &
$\delta\ell[g(\eta)]$ & $\quad\delta\ell_\Pi\quad$ & $\delta\ell_{\rm lensing}$ & $\qquad\quad \delta \ell_p $  \\
\hline
{\rm 1st} & \phantom{0}3 & 12& $-30$ &  $15$ 		& $-4\phantom{.}$ & 15 & 11 & 0 &  $11 = \phantom{1}151 -\phantom{1}140$  \\
{\rm 2nd} & 10		 & 16& $-17$ &  $11$ 		& $19$ 		  & 19 & 58 & 0 &  $58 = \phantom{1}453 -\phantom{1}395$  \\
{\rm 3rd} & 14		 & 15& $-23$ &  $\phantom{1}6$ 	& $34$		  & 21 & 67 & 0 &  $67 = \phantom{1}755 -\phantom{1}688$  \\
{\rm 4th} & 16		 & 14& $-31$ &  $\phantom{1}4$ 	& $43$		  & 20 & 66 & 1 &  $67 = 1057 - \phantom{1}990$  \\
{\rm 5th} & 17		 & 13& $-36$ &  $\phantom{1}3$ 	& $46$		  & 16 & 59 & 1 &  $60 = 1359 - 1299$ \\
{\rm 6th} & 17		 & 12& $-39$ &  $\phantom{1}2$ 	& $48$		  & 11 & 51 & 2 &  $53 = 1661 - 1608$ \\
\hline
\end{tabular}
\end{table*}

\begin{table*}
\caption{\label{table:te} The shift of the $p$-th peak in  the  power spectrum $\Dlc$ is defined by $\delta\ell_p \equiv 151 (p+0.5) -\ell_p($TE$)$, and notations used here are similar to those in Table \ref{table:tt} .}
\begin{tabular}{c | c  c  c  c |  c  c c | l}
\hline \hline
{\rm $p$-th peak}  & $\delta\ell[\phi_{\rm tot}]$ & $ \delta\ell[ k^2P(k)]$ & $\delta\ell[j_\ell]$ &
$\delta\ell[g(\eta)]$ & $\delta\ell_{\rm monopole} $ &
$\delta\ell_{\rm non-monopole }$ & $\delta\ell_{\rm lensing}$ & $\qquad\quad \delta\ell_p $\\
\hline
\phantom{0}1st & 32 & 11\phantom{0} 	& 21 & $\phantom{-}7$  		  & 71  & $\phantom{-}4 $  & 0 & $75 = \phantom{1}227-\phantom{1}152$\\
\phantom{0}2nd & 27 &  8		& 28 & $\phantom{-}8$  		  & 71  & $-2$  	   & 0 & $69 = \phantom{1}378-\phantom{1}309$\\
\phantom{0}3rd & 23 &  5 		& 28 & $\phantom{-}11\phantom{1}$ & 67  & $-6$  	   & 0 & $61 = \phantom{1}529-\phantom{1}468$\\
\phantom{0}4th & 32 &  4 		& 41 & $\phantom{-}10\phantom{1}$ & 87  & $-3$		   & 1 & $85 = \phantom{1}680-\phantom{1}595$\\
\phantom{0}5th & 28 &  4 		& 40 & $\phantom{-}13\phantom{1}$ & 85  & $-3$ 		   & 1 & $83 = \phantom{1}831-\phantom{1}748$\\
\phantom{0}6th & 19 &  2 		& 41 & $\phantom{-}11\phantom{1}$ & 73  & $-5$		   & 0 & $68 = \phantom{1}982-\phantom{1}914$\\
\phantom{0}7th & 15 &  3 		& 39 & $\phantom{-}9$  		  & 66  & $-5$		   & 0 & $61 = 1133-1072$\\
\phantom{0}8th & 18 &  2 		& 41 & $\phantom{-}8$ 		  & 69  & $-5$		   & 0 & $64 = 1284-1220$\\
\phantom{0}9th & 16 &  2 		& 45 & $\phantom{-}6$ 		  & 69  & $-5$ 		   & 1 & $65 = 1435-1370$\\
10th	       & 11 &  1 		& 34 & $\phantom{-}8$ 		  & 54  & $-6$ 		   & 0 & $48 = 1586-1538$\\
11th	       & 11 &  1 		& 41 & $-4$			  & 49  & $-5$ 		   & 0 & $44 = 1737-1693$\\
12th	       & 13 &  0 		& 28 & $-6$ 			  & 45  & $-7$ 		   & 0 & $38 = 1888-1850$\\
\hline
\end{tabular}
\end{table*}

In the baseline model, we simplify the monopole to be purely cosine,
$\TP\propto\cos(kr_s)$, simplify the LSS to be infinitely thin,
i.e. $g(\eta) = \delta(\eta-\eta_\star)$, and simplify the projection from $k$
modes to $\ell$ modes as one-to-one, $\ell=k(\eta_0-\eta_\star)$. In fact,
all above simplifications are not exactly correct. The phase shift
$\phi_{\rm tot}$ of multipoles $\Theta_\ell$, the finite width of the LSS and
the fact that the projection from $k$ modes to $\ell$ modes is not one-to-one,
introduce peak shifts to the spectra. In addition, dipole $\Theta_1$,
polarization $\Pi$ and ISW effect also contribute a
sub-dominant part to the power spectra. Taking TT as an example, we define the
total peak shift relative to the prediction of the baseline model,
$\delta\ell_p \equiv 302\ p- \ell_p({\rm TT})$, where $\ell_p({\rm TT})$ is the
location of $p$-th peak in the theoretical temperature power spectrum $\Dl$ of
the fiducial cosmology (Fig. \ref{fig:baseline}). In the remainder of this
section, we shall investigate each correction contributing to the total peak
shift $\delta\ell_p$ individually. Note that a positive $\delta\ell_p$ denotes
a shift to smaller $\ell$.

\subsubsection{Phase Shifts in Sources: $\delta\ell[\phi_{\rm tot}]$}

The phase shift $\phi_{\rm tot}$ in the monopole $\TP$  at the LSS induces a peak shift in TT,
$\delta\ell[\phi_{\rm tot}] = \phi_{\rm tot}/\theta_\star$, which is decomposed into three components
$\delta\ell[\phin]+\delta\ell[\phig]+\delta\ell[\phi_{\rm dcp}]$ (Table \ref{table:tt} and Fig. \ref{fig:atstar}).
Similar analysis is also done for EE (Table \ref{table:ee} and Fig. \ref{fig:atstar}) and TE (Table \ref{table:te}).
We choose not to do the decomposition for TE, whose source $\TP\times\Pi$ is not an independent quantity.

\subsubsection{Primordial Power Spectrum: $\delta\ell[k^2P(k)]$}

Each $k$ mode carries different amount of power which is specified by the primordial power spectrum $P(k)$,
a detail not included in the baseline model.
In fact, the temperature power spectrum $\Dl$ is modulated by the primordial power spectrum $P(k)$ as follows,
\be
\Dl \simeq k^2 P(k) \sim \frac{1}{\ell},
\ee
where we have used the scale-invariant primordial power spectrum  $k^2P(k)\sim k^{-1}$,
and the simplified correspondence $\ell = k(\eta_0-\eta_\star)$.

The modulation of the primordial power can be derived in a more rigorous way.
Using Limber approximation \citep{Limber, Loverde2008,Lesgourgues2014},
\bea
\label{eq:deltaj}
\int_0^\infty dx f(x) j_\ell(x) \simeq f(L)\sqrt{\frac{\pi}{2L}},
\eea
where $L = \ell +1/2$, the transfer function is written as
\be
\label{eq:deltabessel}
\Delta_\ell(k) = \frac{1}{k}\sqrt{\frac{\pi}{2L}} g(\eta) \l[\frac{\Theta_0(k,\eta)+\Phi(k,\eta)}{\Phi(k,0)}\r]\Bigg|_{k(\eta_0-\eta) = L} ,
\ee
and the temperature power spectrum is simplified as
\bea
\Dl &&  = \int_0^{\eta_0} d\eta \ k^2 P(k) \ g^2(\eta) \nn\\
&&\times \l[\frac{\Theta_0(k,\eta)+\Phi(k,\eta)}{\Phi(k,0)}\r]^2\Bigg|_{k=  L/(\eta_0-\eta)},
\eea
where we have used the definition (\ref{eq:tps}), and changed the integration variable from $k$ to $\eta$.
Taking a step further, using facts that the visibility function is narrowly peaked
at $\eta_\star$ and  $\TP(k,\eta_\star)/\Phi(k,0) \propto \cos(kr_{s,\star}+\phi_{\rm tot})$,
we have
\be
\label{eq:deltajg}
\Dl[j=\delta,g=\delta] \propto \frac{\cos^2(\ell \theta_\star+\phi_{\rm tot})}{\ell},
\ee
where  $\Dl[j=\delta,g=\delta]$  denotes the approximate power spectrum
calculated using the Limber approximation and instantaneous recombination.

The modulation by the primordial power $k^2P(k)\sim 1/\ell$ drives the TT peaks
to smaller $\ell$ from the predictions of the baseline model.
Analytically, it is straightforward to obtain
\be
\delta\ell[k^2P(k)] \simeq \frac{1}{2\theta_\star(p\pi-\phi_{\rm tot})} \simeq \frac{48}{p\pi-\phi_{\rm tot}},
\ee
which is consistent with numerical results (see Table \ref{table:tt}). Similar analysis is also done for EE and TE.

\subsubsection{Asymmetric Projection: $\delta\ell[j_\ell]$}

Assuming the monopole $\TP(k,\eta_\star)$ peaks at $k_p$,
the baseline model predict a peak in the power spectrum at $\ell_p = k_p(\eta_0-\eta_\star)$.
But the projection from $k$ modes to $\ell$ modes is not one-to-one, instead all $k$ modes $k\gtrsim \ell/(\eta_0-\eta_\star)$ contribute to $\ell$, and modes $k\lesssim \ell/(\eta_0-\eta_\star)$ contribute no power to  $\ell$ (Fig.\ref{fig:proj} and Eq.(\ref{eq:bessel})). As a result, a slightly smaller $\ell$ (than $\ell_p$) receives power from a wider range of $k$ modes around $k_p$. Therefore the asymmetric projection drives TT peaks to smaller $\ell$ from  the baseline model predicted peak locations.

The $k$ modes and $\ell$ modes are connected by the transfer function $\Delta_\ell(k)$ [Eq.(\ref{eq:trsf})],
which is simplified as
\be
\label{eq:deltag}
\Delta_\ell(k) \simeq j_\ell(k(\eta_0-\eta_\star))\times \left[\frac{\Theta_0(k,\eta_\star)+\Phi(k,\eta_\star)}{\Phi(k,0)}\right],
\ee
under the approximation of $g(\eta) =\delta(\eta-\eta_\star)$.
Quantitatively, we compute an approximate spectrum $\Dl[g=\delta]$ from Eqs.(\ref{eq:tps}, \ref{eq:deltag}), and numerically measure $\delta\ell[j_\ell]$ from differences between the peak locations of $\Dl[j=\delta, g=\delta]$ and $\Dl[g=\delta]$ (see Table \ref{table:tt}). Similar analysis is also done for EE and TE power.

According to Table \ref{table:tt}, \ref{table:ee} and \ref{table:te}, the asymmetric projection drives the peaks in the spectra to smaller $\ell$ except the first EE peak. The anomaly also comes from the rise in the amplitude of $\Pi$ as $k$ increases. The first peak  in $\Pi$ is tiny compared to following few peaks. As a result, the first EE peak gains more power from larger $k$ modes of $\Pi$, therefore is driven to larger $\ell$ (negative $\delta\ell$).

\subsubsection{Visibility Function: $\delta\ell[g(\eta)]$}

\begin{figure}
\includegraphics[scale=0.42]{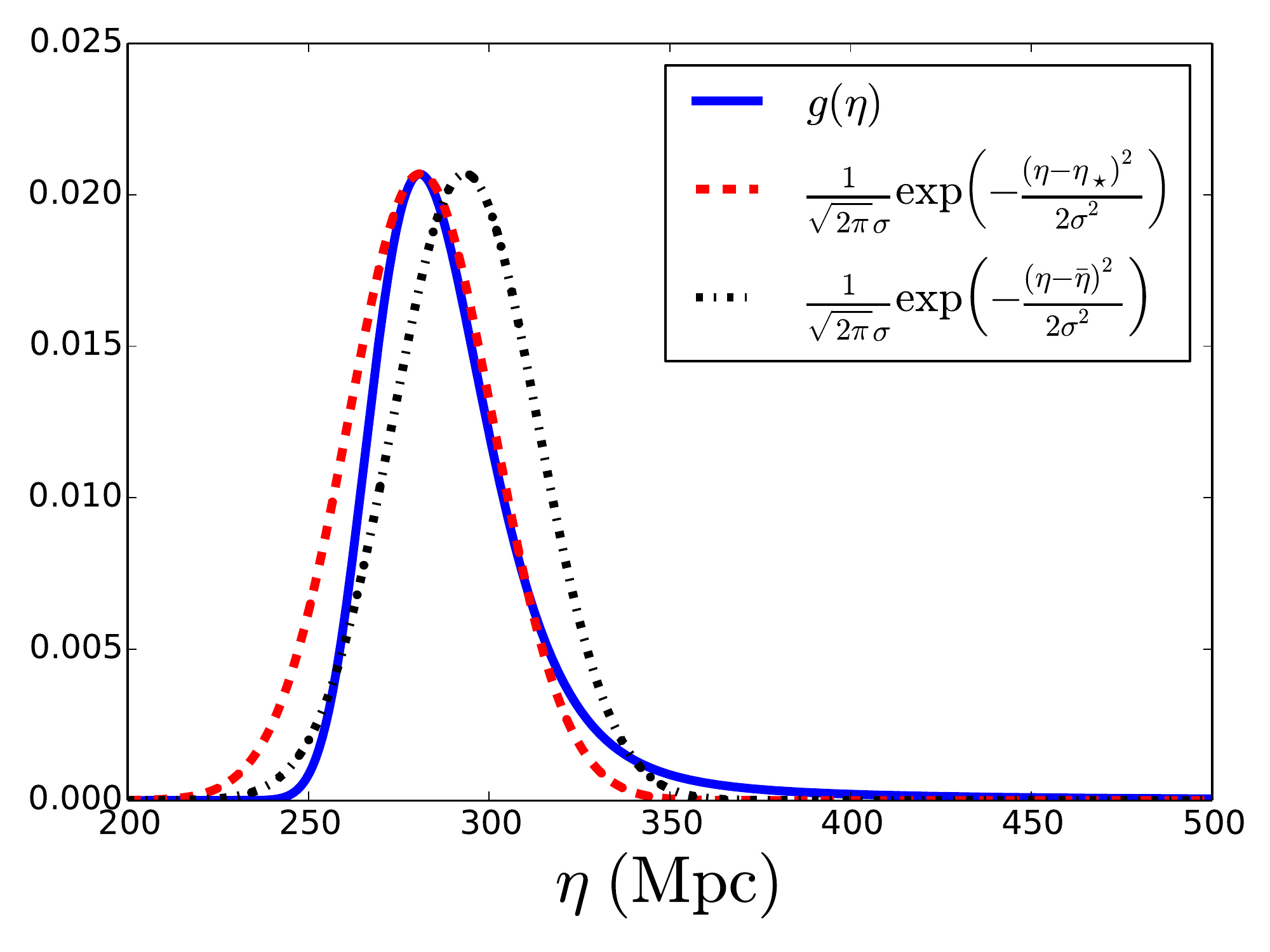}
\caption{\label{fig:visb} The comparison between the asymmetric visibility function $g(\eta)$
and two Gaussian functions both with $\sigma = 20 $ Mpc, peaking at $\eta_\star = 281$ Mpc and $\bar\eta = 293$ Mpc, respectively,
where $\bar\eta$ is the mean decoupling time defined by $\bar\eta = \int g(\eta) \eta \ d\eta/\int g(\eta)\  d\eta$.}
\end{figure}

Due to the finite width of the visibility function $g(\eta)$,
$\Dl$ is powered by the source $g(\eta) [\Theta_0(k,\eta)+\Phi(k,\eta)]$ from a time interval instead of a single time slice $\eta_\star$.
The visibility function  $g(\eta)$ is positively skewed (Fig. \ref{fig:visb}).
But $\TP(k,\eta)$ decreases in amplitude over time due to diffusion damping, so is negatively skewed.
In addition, the monopole $\TP(k,\eta)\propto\cos(kr_s(\eta)+\phi_{\rm tot})$ peaks at smaller $k$ modes at later time (larger $r_s(\eta)$).
Therefore the first few TT peaks  would be driven to smaller $\ell$, where the damping is small and the asymmetry of $g(\eta)$ dominates;
and other TT peaks would be driven to large $\ell$, where the asymmetry of $\TP(k,\eta)$ dominates due to stronger damping.

For polarization, the source term $\Pi$ increases in amplitude over time, generated by free streaming.
Consequently, $\Pi(\eta)$ has similar asymmetry to the visibility $g(\eta)$, and so EE and TE peaks are driven to even smaller $\ell$.

Quantitatively,  the phase shift induced by the visibility function
$\delta\ell[g(\eta)]$ is obtained by the differences between peak locations of
true $\Dl$ and of $\Dl[g=\delta]$. The numerical results are consistent with
our qualitative analysis  (Table \ref{table:tt}, \ref{table:ee} and \ref{table:te}).

\subsubsection{ISW, Dipole, Polarization: $\delta\ell[{\rm non-monopole}]$}

\begin{figure}
\includegraphics[scale=0.42]{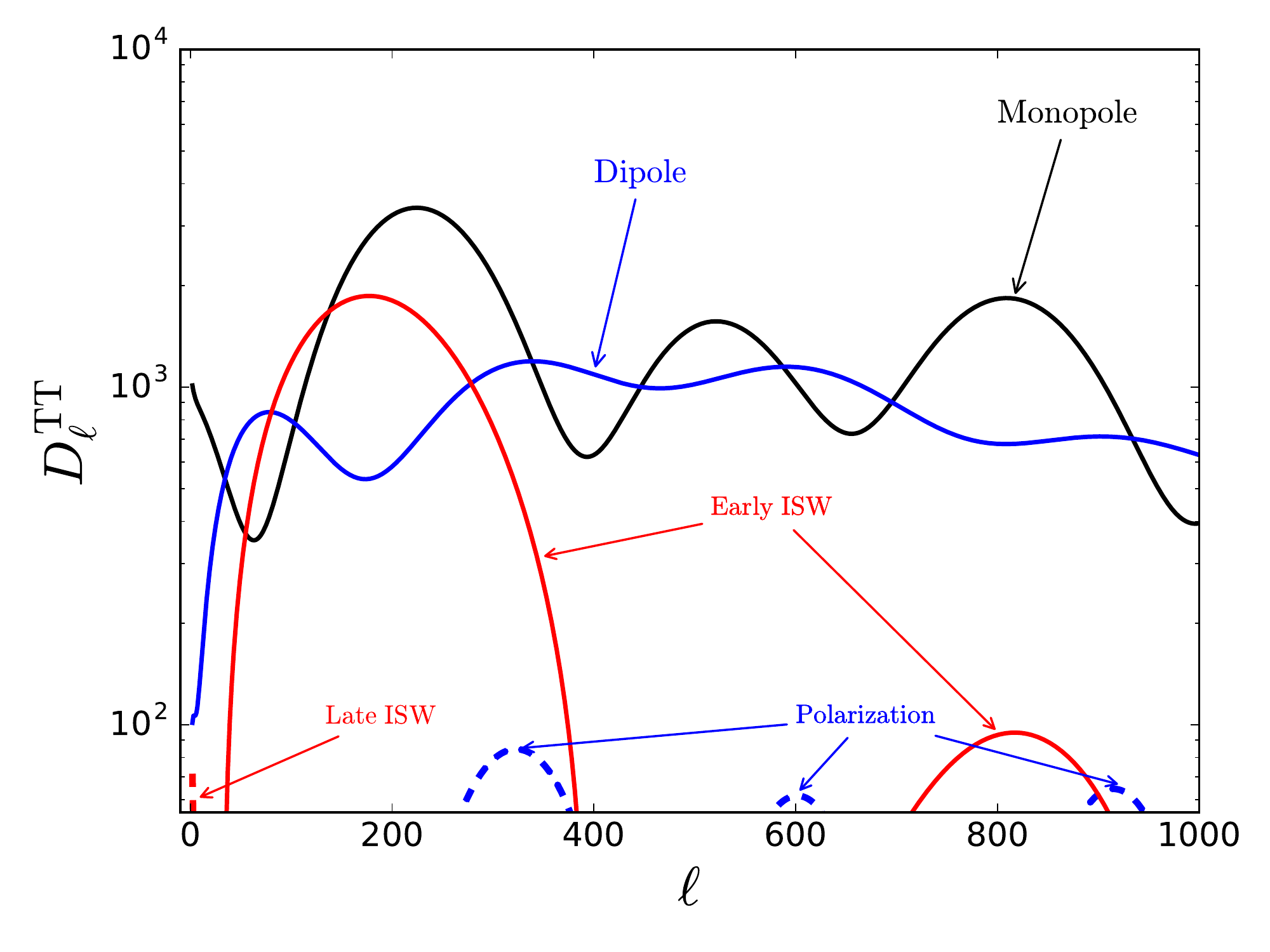}
\caption{\label{fig:nonmono} The contribution to the unlensed $\Dl$ from each
component: dominant monopole, subdominant dipole and early ISW, and negligible polarization and late ISW.}
\end{figure}

The TT and TE power spectra also receive contributions from other components:
dipole $\Theta_1$, polarization source $\Pi$, and ISW effect $\dot\Phi+\dot\Psi$
(see Fig.~\ref{fig:nonmono} for the decomposition of TT).

The early ISW power peaks near the particle horizon scale of decoupling, which
is larger than the sound horizon so drives the first peak to lower $\ell$.
The late ISW only operates at very large scale ($\ell\lesssim 10$),
therefore has almost no impact on the peak locations \citep{Hu1995b}.

According to the baseline model,  $\Theta_1$ is expected to be 90 degrees out
of phase with $\Theta_0$, in which case $\Theta_1$ would have no influence on
the peak locations. Taking the decoupling effect into account, $\Theta_1$ and
$\Theta_0$ are found to be out of phase by more than 90 degrees
(see Appendix \ref{app:dcp}). As a result, $\Theta_1$ drives the peaks to
larger $\ell$. The impact of  $\Pi$ is negligible.

To summarize, $\delta\ell[{\rm non-monopole}]$ of TT and TE are positive at
small $\ell$ modes where the (early) ISW dominates, and are negative at large
$\ell$ modes where $\Theta_1$ dominates (see Fig.~\ref{fig:nonmono}, Table \ref{table:tt} and \ref{table:te}).

\subsubsection{Lensing: $\delta\ell[{\rm lensing}]$ }

Gravitational lensing tends to smooth the power spectra by redistributing the power among $\ell$ modes (see the review of weak lensing by \citep{Lewis2006}). The net effect is that peaks lose power and troughs gain power. If a peak is symmetric, the modes on both sides of the peak would lose the same amount of power, thus the peak amplitude is suppressed with the peak location unaffected. If a peak is asymmetric due to more damping at larger $\ell$ modes, the modes on the right side of the peak would lose more power than modes on the left side, therefore the peak is driven to smaller $\ell$. The EE and TE peaks are more symmetric than TT peaks; as a result, the lensing driven phase shifts in EE and TE are smaller (see Table \ref{table:tt} and Table \ref{table:ee}, \ref{table:te}  for comparison) .

\section{Comparison of predicted and measured peak locations}
\label{sec:measure}

\begin{table}
\label{table:peaks}
\caption{Locations of the peaks in the power spectra. The peak locations
measured from the {\Planck} 2015 data are listed in the 3rd column
\citep[Table E.2. in][]{PlanckXIII.2015}, and the peak locations predicated
by the fiducial cosmology  are listed in the 2nd column (Note that these peak
locations are determined by the fitting procedure used on the data, therefore
are different from the literal peak locations of theoretical power  spectra). }
\begin{tabular}{c c c}
\hline \hline
{\rm $p$-th peak}  & multipole (model) & \phantom{xx} multipole (data)  \\
\hline
\\
$\mathbf{TT}$ {\bf power spectrum} \\
\\
{\rm 1st} & $220.9$ 		& $\phantom{1}220.0\pm0.5$ \\
{\rm 2nd} & $538.5$ 		& $\phantom{1}537.5\pm0.7$ \\
{\rm 3rd} & $809.5$ 		& $\phantom{1}810.8\pm0.7$ \\
{\rm 4th} & $1122.5\phantom{1}$ & $1120.9\pm1.0$ \\
{\rm 5th} & $1445.7\phantom{1}$ & $1444.2\pm1.1$ \\
{\rm 6th} & $1774\phantom{.11}$ & $1776 \pm 5$ \\
{\rm 7th} & $2071\phantom{.11}$ & $\phantom{1}2081\pm25$\\
{\rm 8th} & $2429\phantom{.11}$ & $\phantom{1}2395\pm24$ \\
\\
$\mathbf{EE}$ {\bf power spectrum} \\
\\
1st & $135.3$ 		  & $\phantom{1}137\pm 6$\\
2nd & $395.4$ 		  & $\phantom{1}397.2\pm0.5$ \\
3rd & $689.7$ 		  & $\phantom{1}690.8\pm0.6$ \\
4th & $992.1$ 		  & $\phantom{1}992.1\pm1.3$\\
5th & $1299\phantom{.11}$ & $1296\pm4$ \\
\\
$\mathbf{TE}$ {\bf power spectrum} \\
\\
1st 		& $149.0$ 		& $\phantom{1}150.0\pm0.8$ \\
2nd 		& $307.8$ 		& $\phantom{1}308.5\pm0.4$ \\
3rd 		& $471.3$ 		& $\phantom{1}471.2\pm0.4$ \\
4th 		& $593.4$ 		& $\phantom{1}595.3\pm0.7$ \\
5th 		& $747.3$ 		& $\phantom{1}746.7\pm0.6$ \\
6th 		& $915.9$ 		& $\phantom{1}916.9\pm0.5$\\
7th 		& $1073.3\phantom{1}$   & $1070.4\pm1.0$\\
8th 		& $1224.8\phantom{1}$   & $1224.0\pm1.0$\\
9th 		& $1371.9\phantom{1}$   & $1371.7\pm1.2$\\
10th\phantom{1} & $1542.1\phantom{1}$   & $1536.0\pm2.8$\\
11th\phantom{1} & $1700.6\phantom{1}$   & $1693.0\pm3.3$ \\
12th\phantom{1} & $1865\phantom{.11}$   & $1861\pm4$\\
\hline
\end{tabular}
\end{table}

In contrast to theoretical power spectra, it is impossible to directly read the peak locations out of data points in the presence of noise. To measure the peak locations from the data points, a fitting procedure is required. Taking the TT power spectrum as an example, the {\it Planck} collaboration \citep{PlanckXIII.2015} first removed the damping tail, and then fit Gaussian functions to the peaks in $\Dl$. \footnote{Different fitting procedures were also inviestigated in previous works \citep[e.g.][]{Aghamousa2012,Aghamousa2015}.} The peak locations measured in this specific procedure cannot be compared with the literal peak locations in the theoretical power spectrum of the fiducial cosmology.  To compare the $\Lambda$CDM model predcitons with the peak location measurements , we apply the same fitting procedure on the theoretical spectrum of the  fiducial cosmology (Table \ref{table:peaks}). We find that peak locations measured from the data points and from the theoretical spectra are in agreement.  Most of the relative displacements are within $1\sigma$, and all the relative displacements are less than $3\sigma$.

\section{Conclusions}
\label{sec:conclusion}

The acoustic peak locations of the angular power spectra have been studied in
detail. We start from a baseline model, which assumes tight coupling between
photons and baryons before instantaneous recombination, and  a simplified
one-to-one projection from $k$ modes of the photon perturbations at the LSS to
the $\ell$ modes of  the angular power spectra. Taking the temperature power
spectrum as an example, the baseline model predict that
$\TP(k,\eta) \propto \cos(kr_s(\eta))$ and
$\Dl \propto  \cos^2(kr_{s,\star}) |_ {\ell = k(\eta_0-\eta_\star)}= \cos^2(\ell \theta_\star)$,
which peaks at $\ell_p^0 = p \pi/\theta_\star = 302\ p$. The baseline model is
roughly correct in its prediction for the peak spacing and for the relative
positions of the peaks in different spectra, but is off by a large margin in
its absolute predictions of peak locations. For example, the first peak in
$\Dl$ is at $\ell_1 = 220$, which is shifted  by $\delta\ell_1 = 82$ from the
baseline model prediction $\ell_1^0=302$. The shift of the true power spectra
locations relative to the baseline model predictions comes both from the phase
shift  $\phi_{\rm tot}$ in the acoustic oscillations of the photon perturbations
$\TP(k,\eta) \propto \cos(kr_s(\eta)+\phi_{\rm tot})$, and the fact that the
projection from photon perturbations at the LSS to the angular power spectra is
far more complicated than assumed in the baseline model.

The phase shift $\phi_{\rm tot}(k, \eta_\star)$ consists of two components
$\phi_{\rm dcp}$ and $\phi_{\rm gr}$, where $\phi_{\rm dcp}$ is the phase shift
induced by decoupling and dominates for large $k$ modes ($k\eta_\star\gtrsim 1$),
and $\phi_{\rm gr}$ is the phase shift induced by the gravitational driving and
dominates for small $k$ modes ($k\eta_\star\lesssim 1$). The latter component
can be further decomposed as $\phig+\phin$, where $\phig$ is the transient
induced phase shift and  $\phin$ is the neutrino induced phase shift. A key
difference between the two is that $\phig$  decays with increasing $k$, while
$\phin$ grows with increasing $k$ and approaches a nonzero constant. This
difference stems from the time dependence of these effects, and the fact that at
higher $k$ there is more time, in units of the natural period of the oscillator
$2\pi/(kr_s)$, between horizon-crossing and matter-radiation equality.

The projection from $k$ modes to $\ell$ modes is not one-to-one as assumed in
the baseline model. All perturbation modes satisfying $k \gtrsim \ell/(\eta_0-\eta_\star)$
contribute to the angular power spectra at a given $\ell$.  In addition, the LSS
has non-zero width. Both of these differences with the baseline model introduce
peak shifts to the power spectra. Other effects including the modulation of the
primordial power spectrum $P(k)$, (early) ISW effect, dipole moment of photon
perturbation and lensing also contribute subdominant shifts to the peak locations.

We also compare each peak location determined from \Planck measurements to the
location predicted under the assumption of the
best-fit $\Lambda$CDM model, and find consistency.

Our entire motivation in pursuing this work was to achieve an understanding of
the numerically calculated predictions of the
$\Lambda$CDM model. However, we now
speculate on a potential application. With further development, perhaps our
analytic understanding of the shifts in
the peak locations could be combined
with the morphing procedure of \citet{Sigurdson2000}, to achieve a highly accurate, very fast, Boltzmann code emulator,
improving upon interpolative schemes such as PICO \citep{Fendt2007}.

\section{acknowledgements}
We thank the anonymous referee for his/her valuable comments and suggestions,
including the suggestion that there might be an application to Boltzmann code emulation.
We also thank Douglas Scott for careful reading this manuscript and introducing the history of first TT peak detection.
This work made extensive use of the NASA Astrophysics Data System and of the {\tt astro-ph} preprint archive at {\tt arXiv.org}.
All the numerical calculations are done with the Boltzmann  code {\tt Class}.

\appendix

\section{$\phi_{\lowercase{\rm dcp}}$}
\label{app:dcp}

High-order corrections to the tight coupling approximation are only important at late time after a  mode enters the horizon ($k\eta \gtrsim 1$), and the gravitational potentials have decayed. Following \citet{Zaldarriaga1995}, we set $\Phi=0$ and $\Psi = 0$ in this subsection (Refer to \citet{Blas2011} for rigorous high-order correction to the tight coupling approximation).
Assuming the formal solutions
\be
\Theta_i, \Theta_{p,i}, V_b \sim e^{i \int \omega d\eta},
\ee
the dipole moment of Eq.(\ref{eq:EB1}) is written as
\be
\label{eq:dipole}
\Theta_1 - \frac{i}{3}V_b = \frac{1}{\dt} \l[i\omega\Theta_1 + k \l( \frac{2}{3}\Theta_2 - \frac{1}{3}\Theta_0\r) \r].
\ee
Accurate to $O(k/\dot \tau)^3$ on both sides of the above equation, Eqs.(\ref{eq:EB1}, \ref{eq:EB2}) are decomposed as
\bea
\label{eq:multipole1}
\Theta_0 &=& \frac{ik}{\omega}\Theta_1, \\
\Theta_2 &=& -\frac{8}{15}\frac{k}{\dt} \Theta_1 \l(1+ \frac{11}{6}\frac{i\omega}{\dt} \r) ,\nn\\
\Pi &=& \frac{5}{2} \Theta_2 \l( 1 + \frac{3}{2}\frac{i\omega}{\dt} \r), \nn
\eea
and Eq.(\ref{eq:EB3}) is expanded as
\be
\label{eq:multipole2}
\Theta_1 - \frac{i}{3}V_b = \Theta_1\l[-i\frac{\omega R_b}{\dt} + \l(\frac{\omega R_b}{\dt}\r)^2 + i\l(\frac{\omega R_b}{\dt}\r)^3  \r],\quad\
\ee
where we have dropped a term $\frac{R_b}{\dt}\frac{\dot a}{a}$ in the bracket
on the right-hand side, which is smaller than $\frac{R_b}{\dt}\omega$
when the mode is within the horizon $k\eta >  1$.

Plugging Eqs.(\ref{eq:multipole1},\ref{eq:multipole2}) into Eq.(\ref{eq:dipole}),
we obtain, for $\omega = \omega_0 + \delta\omega_0 + i\gamma$,
\bea
\omega_0 &=& k c_s, \quad \frac{\gamma}{\omega_0} = -\frac{k}{\dt}\frac{\l(c_s^2 R_b^2 + \frac{16}{45} \r)}{2c_s(1+R_b)},\\
\frac{\delta\omega_0}{\omega_0} &=&
\frac{k^2}{\dt^2} \frac{\l( c_s^2 R_b^2 +\frac{88}{135} \r) -\frac{3}{4}\l(c_s^2 R_b^2 +\frac{16}{45}\r) \l(5c_s^2 R_b^2 + \frac{16}{45}\r) }{2  (1+R_b)} . \nn
\eea
Note that $\dt$ is negative, and so $\gamma$ is positive.
Therefore, accurate to $O(k/\dt)^3$, $\Theta_0(k,\eta)$ can be described
as a damped oscillator with a time-dependent phase shift, i.e.
\be
\Theta_0(k, \eta) \propto \cos(kr_s + \phi_{\rm dcp}) e^{-k^2/k_D^2} ,
\ee
where
\bea
\label{eq:dcp}
\phi_{\rm dcp} &=& \int\delta\omega_0 \  d\eta \sim O(k/\dt)^2,  \nn\\
 k^2/k_D^2 &=& \int \gamma \ d\eta \sim O(k/\dt).
\eea

Other useful multipoles obtained are
\bea
\Theta_1 &\propto&  c_s \sin(kr_s + \phi_{\rm dcp} + \phi_1)e^{-k^2/k_D^2}, \nn\\
\Pi &\propto& -\frac{4k}{3\dt}c_s\sin\l( kr_s + \phi_{\rm dcp} + \phi_2\r) e^{-k^2/k_D^2},
\eea
where
\bea
\phi_1 = \tan^{-1}\l(\frac{\gamma}{\omega_0} \r), \quad \phi_2 = -\tan^{-1}\l(\frac{21\gamma}{4\omega_0}\r),
\eea
and $\phi_1 (\phi_2)$ comes from the fact that $\Theta_1 (\Pi)$ and $\Theta_0$
are not exactly $90$ degrees out of phase due to the diffusion damping.

\section{$\phig$}
\label{app:phig}

The solution to the equation of the forced oscillator [Eq.(\ref{eq:driving})] is
written as (e.g. Eq. (8.24) of \citet{Dodelson03})
\bea
\dg(kr_s)
&&= \dg(0)\cos(kr_s) \nn\\
&&- 3\int_0^{kr_s} d(kr_s') \Phi_+(kr_s') \sin(kr_s - kr_s'),
\eea
which can be simplified as
\bea
&&\dg(kr_s)  \nn\\
&=& \l[ \dg(0) + 3 A(kr_s) \r] \cos(kr_s)  - 3 B(kr_s) \sin(kr_s), \nn\\
&=& \sqrt{\l[ \dg(0) + 3 A \r]^2 + (3B)^2} \cos(kr_s +\theta),
\eea
where $\dg(0)$ is the initial amplitude denoted as $\dg(0) \equiv -3\zeta$, and
\bea
\label{eq:AB}
A(kr_s) &=& \int_0^{kr_s} \Phi_+(kr_s') \sin(kr_s') \ d(kr_s') ,\nn\\
B(kr_s) &=& \int_0^{kr_s} \Phi_+(kr_s') \cos(kr_s') \ d(kr_s') ,
\eea
\be
\label{eq:shift}
\theta = \sin^{-1}\l( \frac{3B}{\sqrt{(3A + \dg(0))^2 + (3B)^2} } \r).
\ee

To solve $A$ and $B$, $\Phi_+$ is required.
In the radiation domination, $\Phi_+$ is sourced by $\Phi_-$  (e.g. Eq.(2.50) of \citet{Baumann2016})
\be
\label{eq:phip}
\Phi''_+ + \frac{4}{kr_s}\Phi'_+ + \Phi_+= S(\Phi_-) \equiv \Phi''_- + \frac{2}{kr_s}\Phi'_- + 3 \Phi_- ,
\ee
and $\Phi_-$ is sourced by the radiation stress which is dominated by free-streaming neutrinos
(e.g. Eq.(5.33) of \citet{Dodelson03})
\be
\label{eq:stress}
k^2 \Phi_- = -32\pi Ga^2 \rho_\nu\mathcal{N}_2 ,
\ee
where we have dropped the negligible stress of photons.

Assuming a cosmology without neutrinos, or accurate to $O(R_\nu^0)$, where $R_\nu$
is the energy fraction of neutrinos in radiation, $R_\nu \equiv \rho_{\nu}/(\rho_\nu+\rho_\gamma)$,
we have $\Phi_-^{(0)}=0$, and $\Phi_+^{(0)}$ can be analytically solved as
\be
\label{eq:phip0}
\Phi_+^{(0)}(kr_s) = -4\zeta\frac{\sin (kr_s) - kr_s \cos(kr_s)}{(kr_s)^3}.
\ee
Consequently, $A$ and $B$ are integrated as
\bea
A^{(0)}(kr_s) &=& 2\zeta\l(1-\frac{\sin^2(kr_s)}{(kr_s)^2} \r)\xrightarrow{\scriptscriptstyle kr_s\to\infty} 2\zeta, \nn\\
B^{(0)}(kr_s) &=& 2\zeta\frac{kr_s - \cos(kr_s)\sin(kr_s)}{(kr_s)^2}\xrightarrow{\scriptscriptstyle kr_s\to\infty}0.
\eea
Plugging $A^{(0)}$ and $B^{(0)}$ into Eq.(\ref{eq:shift}), we obtain the phase shift $\theta$ for $\TS$.
With $\dg$ and $\Phi_{\pm}^{(0)}$ known, it is straightforward to obtain both $\TP$ and its phase shift $\phig$,
which is  indistinguishable from $\theta$ after the mode enters the horizon.

\section{$\phin$}
\label{app:phin}
Following \citet{Baumann2016}, $\phin$ can be analytically studied by a perturbation approach,
whose earlier version was orginally developed for probing the impact of neutrino free-streaming
on tensor modes \citep[e.g.][]{Weinberg2004, Dicus2005, Watanabe2006, Miao2007, Xia2008}.
Accurate to $O(R_\nu)$, the two potential are written as
\be
\Phi_- = R_\nu \Phi_-^{(1)}, \quad \Phi_+ =  \Phi_+^{(0)} +  R_\nu \Phi_+^{(1)} .
\ee

In the radiation domination, Eq.(\ref{eq:stress}) can be rewritten as
\be
\label{eq:phim}
\Phi_- = -\frac{4}{k^2r_s^2} R_\nu \mathcal{N}_2=-\frac{4}{3k^2r_s^2} R_\nu D_{\nu,2},
\ee
where $D_\nu$ is defined by $D_\nu \equiv 3 (\mathcal{N} - \Psi)$, and $ D_{\nu,2}$ is the quadrupole moment, $ D_{\nu,2} = 3 \mathcal{N}_2$.
To determine $\Phi_-$ to $O(R_\nu)$, we only need to specify $D_{\nu,2}$ to $O(R_\nu^0)$.

The evolution of neutrino perturbation is governed by (e.g. Eq.(4.107) of \citet{Dodelson03}),
\be
\dot\Dn + ik\mu \Dn = - 3 ik\mu \Phi_+,
\ee
and the solution is
\bea
\Dn(\eta)
&=& \Dn(0) e^{-ik\mu \eta} \nn\\
&&- 3ik\mu\int_0^\eta d\eta' \ e^{-ik\mu(\eta-\eta')}\Phi_+(\eta'),
\eea
where $D_\nu(0)$ is determined by the inflation inspired initial condition $D_\nu(0) = D_{\nu,0}(0) =  -3 \zeta$.
Plugging the zeroth-order potential $\Phi_+^{(0)}$ into the above solution, $D_\nu(\eta)$ and consequently $D_{\nu,2}$ of $O(R_\nu^0)$ are obtained.

Combining Eq.(\ref{eq:phim}) and Eq.(\ref{eq:phip}), $\Phi_-^{(1)}(kr_s)$ and $\Phi_+^{(1)}(kr_s)$ are obtained. Plugging $\Phi_+^{(1)}(kr_s)$ into Eqs.(\ref{eq:AB}), $A$ and $B$ can be numerically obtained, accurate to $O(R_\nu)$ \citep{Baumann2016},
\bea
A(kr_s\rightarrow\infty) &=& 2\zeta\l(1-0.134 R_\nu \r), \nn\\
B(kr_s\rightarrow\infty) &=& 0.600 \zeta R_\nu.
\eea
Plugging $A$ and $B$ into Eq.(\ref{eq:shift}), the phase shift induced by neutrinos accurate to $O(R_\nu)$ is obtained
\be
\phin = \frac{B}{\zeta} = 0.191 R_\nu \pi.
\ee
The above $\phin$ is derived under the assumption of radiation domination, so is appropriate only for large $k$ modes. Taking the matter domination epoch into account, $\phin$ is expected to approach zero as $\sim k^2$ for small $k$ modes \citep{Baumann2016}.

\bibliographystyle{mnras}
\bibliography{ms}

\end{document}